%% file: main.tex
\documentclass[sigconf, 9pt]{acmart}
\AtBeginDocument{%
 }

\usepackage{algorithm,algpseudocode}
\usepackage{fontawesome} %added for \falock - lock symbol
\usepackage{booktabs} %For table style
\usepackage{graphicx} %For table
\usepackage{multirow} %table multi-row
\usepackage{subcaption}

\newcommand{\HZ}[1]{\textcolor{red}{[HZ: #1]}}

\makeatletter
\newcounter{phase}[algorithm]
\newlength{\phaserulewidth}
\newcommand{\setphaserulewidth}{\setlength{\phaserulewidth}}
\newcommand{\phase}[1]{%
  \vspace{-1.25ex}
  \Statex\leavevmode\llap{\rule{\dimexpr\labelwidth+\labelsep}{\phaserulewidth}}\rule{\linewidth}{\phaserulewidth}
  \Statex\strut\refstepcounter{phase}\textit{\thephase~--~#1}% Phase text
  \vspace{-1.25ex}\Statex\leavevmode\llap{\rule{\dimexpr\labelwidth+\labelsep}{\phaserulewidth}}\rule{\linewidth}{\phaserulewidth}}
\makeatother

\setphaserulewidth{.7pt}

\newcommand{\sys}{\texttt{DREAMPlaceFPGA-MP}}
\begin{document}

\author{Zhili Xiong}
\affiliation{%
  \institution{The University of Texas at Austin}
  % \streetaddress{P.O. Box 1212}
  \city{Austin}
  \state{Texas}
  \country{USA}
}
\email{zhilix691@utexas.edu}

\author{Rachel Selina Rajarathnam}
\affiliation{%
  \institution{The University of Texas at Austin}
  % \streetaddress{P.O. Box 1212}
  \city{Austin}
  \state{Texas}
  \country{USA}
}
\email{rachelselina.r@utexas.edu}

\author{Zhixing Jiang}
\affiliation{%
  \institution{The University of Texas at Austin}
  % \streetaddress{P.O. Box 1212}
  \city{Austin}
  \state{Texas}
  \country{USA}
}
\email{zx.jiang@utexas.edu}

\author{Hanqing Zhu}
\affiliation{%
  \institution{The University of Texas at Austin}
  % \streetaddress{P.O. Box 1212}
  \city{Austin}
  \state{Texas}
  \country{USA}
}
\email{hqzhu@utexas.edu}

\author{David Z. Pan}
\affiliation{%
  \institution{The University of Texas at Austin}
  % \streetaddress{P.O. Box 1212}
  \city{Austin}
  \state{Texas}
  \country{USA}
}
\email{dpan@ece.utexas.edu}

\title{\huge{\texttt{DREAMPlaceFPGA-MP}: An Open-Source GPU-Accelerated Macro Placer for Modern FPGAs with Cascade Shapes and Region Constraints}}


\renewcommand{\shortauthors}{Zhili Xiong, Rachel Selina Rajarathnam, Zhixing Jiang, Hanqing Zhu, \& David Z. Pan}

\settopmatter{printacmref=false} % Removes citation information below abstract
\renewcommand\footnotetextcopyrightpermission[1]{} % removes footnote with conference information in first column
\pagestyle{plain} % removes running headers

\input{texts/Abstract}




\maketitle

\input{texts/Introduction}

\vspace{1em}
\input{texts/Background}
\input{texts/Methodology-GP}
\input{texts/Methodology-LG}
\input{texts/Experiments}

\input{texts/Conclusion}

\begin{acks}
The authors would like to thank the MLCAD 2023 FPGA Macro-Placement Contest organizers, in particular Dr. Ismail Bustany (AMD) for running this industry-strength contest and sharing the results.  
\end{acks}
\balance %Balance bibilography
\bibliographystyle{ACM-Reference-Format}
\bibliography{main}

\end{document}

%% file: texts/Abstract.tex
\begin{abstract}
 
%%%%
FPGA macro placement plays a pivotal role in routability and timing closer to the modern FPGA physical design flow.
In modern FPGAs, macros could be subject to complex cascade shape constraints requiring instances to be placed in consecutive sites.
In addition, in real-world FPGA macro placement scenarios,
designs could have various region constraints that specify boundaries within which certain design instances and macros should be placed.
In this work, we present \sys{}, an open-source GPU-accelerated FPGA macro-placer that efficiently generates legal placements for macros while honoring cascade shape requirements and region constraints.
Treating multiple macros in a cascade shape as a large single instance and restricting instances to their respective regions, \sys{} obtains roughly legal placements.
The macros are legalized in multiple steps to efficiently handle cascade shapes and region constraints.
Our experimental results demonstrate that \sys{} is among the top contestants of the MLCAD 2023 FPGA Macro-Placement Contest.

\end{abstract}

%% file: texts/Introduction.tex
\section{Introduction}\label{sec:intro}
Modern FPGAs incorporate numerous macros, particularly for intellectual property (IP) blocks, in addition to fundamental components like digital signal processors (DSPs), random access memories (RAMs), look-up tables (LUTs), flip-flops (FFs), and input-output (IOs) blocks. 
 These macros are primarily based on DSP and RAM. 
The composition of macros on an FPGA plays a pivotal role in enhancing performance, making them a crucial element in contemporary FPGA architecture designed to meet the performance demands of high-performance applications.
For instance, the Ultrascale+ series ~\cite{XilinxUltraScale}, manufactured using advanced process technology, features various memory macros such as block RAMs (BRAMs) and ultra RAMs (URAMs) tailored for memory-intensive applications. 
It also provides a larger pool of digital signal processors (DSPs) resources for high-performance signal processing, distinguishing it from its predecessor, Ultrascale ~\cite{amd_us}, which had more limited allocations for both macro types and resource quantities.

\iffalse
Modern FPGAs consist of various heterogeneous instances such as digital signal processors (DSPs), random access memories (RAMs), look-up tables (LUTs), flip-flops (FFs), and input-output (IOs) blocks.
While the numerous LUTs and FFs allow for flexibility in programming an FPGA, using intellectual property (IP) blocks such as DSPs and RAMs enables high performance and efficiency.
These large DSP and RAM blocks on an FPGA are called `\textit{macros}.'
The composition of macros on an FPGA continues to increase to meet high-performance application needs.
For instance, an Ultrascale+ architecture consists of three types of macros: DSPs and two types of memories - block RAMs (BRAMs) and ultra RAMs (URAMs), while its predecessor Ultrascale consisted of only DSP and BRAM macro instances~\cite{amd_us}.
\fi

With a considerable number of macros in modern FPGA, the macro placement plays an integral role in routability and timing closure in the physical design flows.
To tackle the FPGA macro placement,
previous placement algorithms generally place macros and other non-macro instances (e.g.,  LUTs and FFs) simultaneously in a \textit{mixed-size manner}.
Simulated annealing-based approaches~\cite{vpr} explicitly ensure legal locations of all instances throughout the placement process.
On the other hand, analytical techniques such as quadratic placement~\cite{UTPlaceF, rippleFPGA, GPlace3.0, HeAP, AMFPlacer} and non-linear placement ~\cite{elfPlace, dreamplacefpga} adopt a two-step approach by placing macros and non-macros first, following by a legalization stage to legalize the roughly legal placement solution.

Recently, the MLCAD 2023 FPGA Macro-Placement Contest~\cite{mlcad2023} introduced updated FPGA macro placement benchmarks, tailored to the Xilinx UltraScale+ architecture with more practical constraints based on real-world applications. 
However, many of the FPGA placement methods mentioned earlier predominantly focused on the ISPD'16 contest benchmarks ~\cite{ispd2016}, which were based on the Xilinx Ultrascale architecture ~\cite{amd_us}. 
ISPD'16 contest benchmarks had a different objective: optimizing the total wirelength for entire designs, and they featured simplified macros with uniform sizes.
Consequently, applying these methods directly to the practical macro placement problem featured in the new benchmark may not yield accurate results.
In real-world FPGA macro placement scenarios, additional complexities come into play, as exemplified by the challenges introduced in the MLCAD 2023 contest benchmark suite.
First, macros can be confined to specific \textit{cascade shapes} based on their connectivity. This entails the precise placement of several macros of the same type in consecutive sites.
Second, macros are constrained to  \textit{regions}, meaning that in the final legalized solution, certain macros can only be placed within specifically designated areas.
Furthermore, the macro placement problem, being of a smaller scale compared to mixed-size placement, necessitates solutions to be delivered within a tighter timeframe.
Hence, the demand for an advanced FPGA macro placement tool is imperative, which is essential to unleashing the potential for generating high-quality macro placement solutions.

In this work, we present \sys{}, an FPGA macro placer that honors cascade shape and region constraints and is built on the open-source FPGA placement framework~\cite{dreamplacefpga}.
The proposed macro placer comprises a global placement stage that considers all instances - macros and non-macros together, followed by a macro legalization stage.
To ensure the order of macro instances in a cascade shape is maintained throughout the placement, we treat a cascade shape as one large instance.
The unified representation of a cascade shape minimizes the displacement to legalize all the macros contained in the cascade shape.
Region constraints are enforced on all instances by restricting them within their designated regions at each global placement iteration.
A three-step macro legalization is employed at the end of global placement to obtain legal macro placements.
The different steps during macro legalization ensure that the cascade shape and region constraints are honored.

Our primary contributions are as follows:
\begin{itemize}
    \item This work presents \sys{}, an open-source macro placer for modern heterogeneous FPGAs with GPU acceleration that honors complex constraints such as cascade shapes and region constraints.
    \item \sys{} employs a unified representation for cascade shapes to ensure the order among the cascade macros is maintained and minimal displacement during legalization.
    \item To ensure region constraints are honored, we restrict all the instances to their respective regions during all the macro placement stages.
    \item During the legalization stage, instead of legalizing all macros together, we perform 3-stage legalization considering the cascade shapes and region constraints.
    \item On the 140 public benchmarks of the MLCAD 2023 FPGA Macro-Placement Contest~\cite{mlcad2023}, \sys{} obtains $5.3\%$, $42.6\%$, and $42.8\%$ lower (the better) overall scores, respectively, compared to the top three contestants. 
    In comparison to Vivado placement, it achieves a $6.32\%$ improvement in placement HPWL and a $11.1\%$ (CPU)/$19.6\%$ (GPU) reduction in total placement runtime.
\end{itemize}

The rest of this paper is organized as follows: 
Section ~\ref{sec:background} provides a foundational view of the considered FPGA architecture and introduces the electrostatics-based placement formulation.
Section ~\ref{sec:methodology} presents our proposed macro placer and details the methodologies across the macro placement stages.
Section ~\ref{sec:exp} presents the results of our experiments, illustrating the effectiveness of the proposed macro placer.
Finally, in Section~\ref{sec:conclusion}, we summarize the work.

%% file: texts/Background.tex
\section{Background}\label{sec:background}
This section provides a background on the considered FPGA architecture and the constraints on macros.
An overview of the electrostatics-based FPGA placement algorithm, employed in the proposed macro placer, is also provided.

%
%%%%%%%%%%%%%%%%%%%
%\vspace{-0.05in}
\subsection{FPGA Architecture}
\begin{figure}[bt]
  \centering
  \includegraphics[width=\columnwidth]{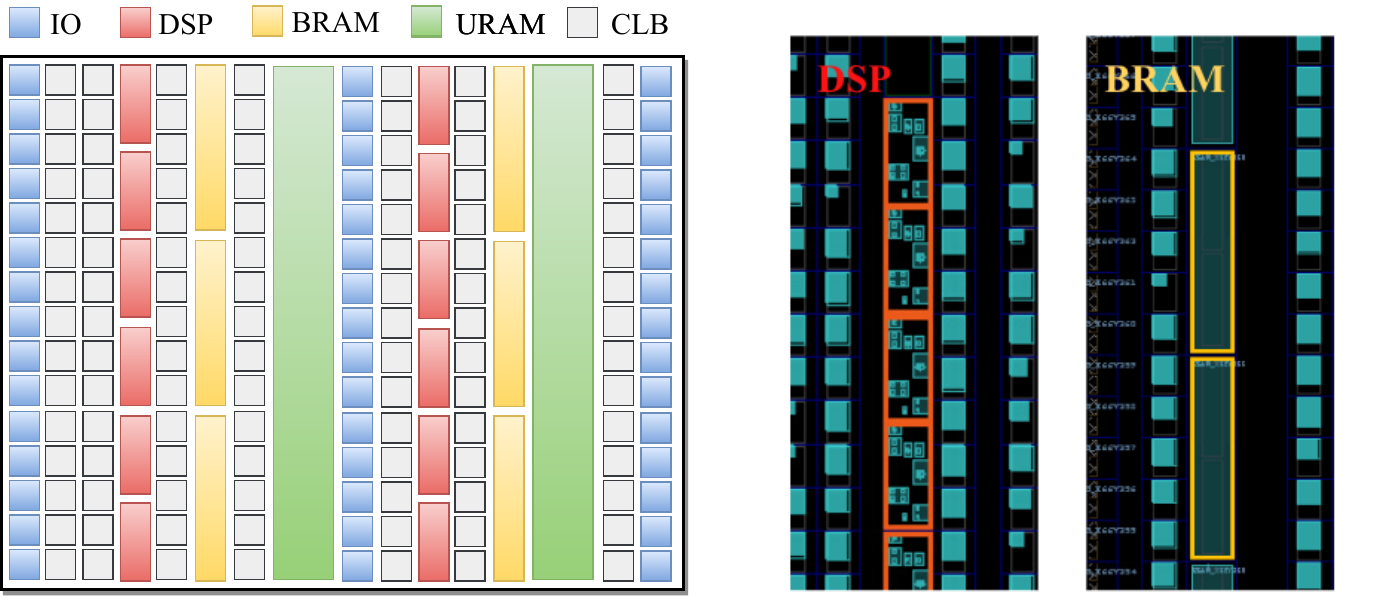}
  %\vspace{-0.1in}
  \caption{Xilinx Ultrascale+ architecture (\texttt{Left}); Cascade shapes in DSPs and BRAMs. (\texttt{Right})}
  %\vspace{-0.1in}
  \Description{The columns representing the macro placement resources.}
  \label{fig:Xilinx_US+}
\end{figure}

This work employs the Xilinx Ultrascale+ architecture~\cite{amd_us} that consists of a column-based layout with heterogeneous site types such as Configurable Logic Blocks (CLBs), Digital Signal Processors (DSPs), memories (BRAMs and URAMs), and Input-Output (IO) blocks, as shown in Figure~\ref{fig:Xilinx_US+}.
A CLB site can accommodate several Look-Up Tables (LUTs), Flip-Flops (FFs), and adders.

The macro instances, such as DSPs, BRAMs, and URAMs, are subject to shape and region constraints.
A \textit{cascade shape} constraint requires multiple macro instances to be placed in continuous site locations in a specified order to ensure proper functionality.
An example of cascade shapes is depicted in Figure~\ref{fig:Xilinx_US+}, where DSPs and BRAMs are grouped to improve overall capacity and function.

\begin{figure}[!t]
    %\vspace{-0.1in}
    \centering
    \includegraphics[width=0.65\columnwidth]{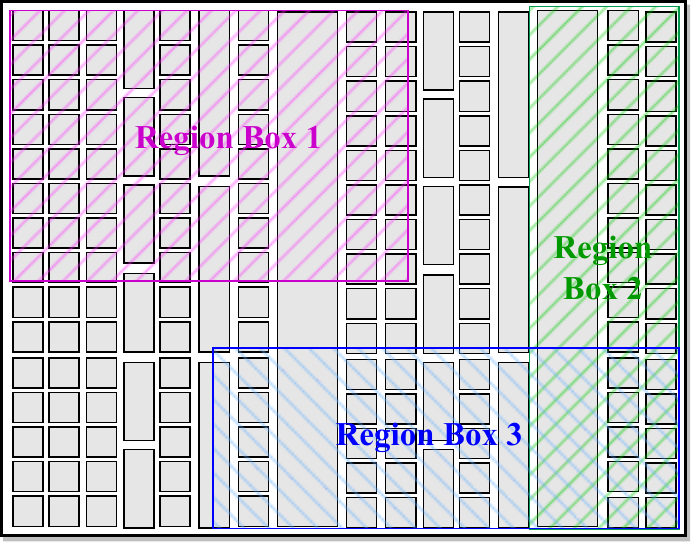}
    \caption{\textit Multiple regions on an FPGA.}
    \vspace{-0.1in}
    \label{fig:Region_Boxes}
    %\vspace{-0.2in}
  
\end{figure}

While cascade shape constraints are limited to macros, any instance can be subject to a region constraint.
As illustrated in Figure~\ref{fig:Region_Boxes}, an FPGA layout can contain multiple regions that could overlap with other regions.
A region can contain several instances and is not limited by the resource type.
The region constraints may originate from clock regions on the chip or can be user-defined.
Instances with region constraints must be placed within their assigned regions to ensure a legal placement.
An instance not assigned a region constraint can be placed on any corresponding FPGA site.
\vspace{1em}

\subsection{Electrostatics-based Placement}
FPGA placement aims to minimize the total wirelength and the overlaps between all the design instances, as shown in Equation~\eqref{eqn:placement_formula} ~\cite{10.1109/TCAD.2008.923063, ePlace, 10.5555/1129601.1129727}.


\begin{equation} \label{eqn:placement_formula}
    \min_{x,y} f(x, y) = \min \left(\sum_{e \in E}W_e(x, y) + \lambda \cdot D(x, y)\right),
\end{equation}

where $E$ represents the set of all the design nets, $W_e$ denotes the half-perimeter wirelength (HPWL) of net $e \in E$, and $D(x,y)$ denotes the density of instances at location $(x, y)$ with lagrangian multiplier $\lambda$.
This work adopts an electrostatic system analogy employed by the state-of-the-art academic FPGA placers~\cite{elfPlace, dreamplacefpga} to represent the density term $D(x,y)$ in Equation~\eqref{eqn:placement_formula}.
The density term is analogous to an electrostatic system's electric potential $\Phi$, where sites are treated as fixed negative charges and design instances are positive charges.
The electric potential gradient is the electric field that moves overlapped instances away from each other toward their respective sites.

To ensure instances of a particular resource type are placed in corresponding sites, each type is solved using a separate electrostatic system. Thus, the placement objective in Equation~\eqref{eqn:placement_formula} can be rewritten using augmented lagrangian relaxation as Equation~\eqref{eqn:GP_formula}, wirelength is calculated using a Weighted Average (WA) model $\widetilde{W}$ ~\cite{5981862,6480847}:
\begin{equation} \label{eqn:GP_formula}
    \min_{x,y} f(x, y) = \widetilde{W}(x, y) + \sum_{s\in S}\lambda_{s}(\Phi_{s}(x, y)+ c_s\Phi_{s}(x, y)^2).
\end{equation}
where $\lambda_{s}$ and $\Phi_{s}$ are the density multiplier and electric potential energy of each resource type, $s \in S$= \{LUT, FF, DSP, BRAM\}, and $c_s$ is a constant that is inversely proportional to the initial potential energy of $s$.
For large values of $\Phi$, the placement formulation aims to minimize the overlaps between the instances emphasized by the quadratic penalty term~\cite{DREAMPlace3.0}. 

%% file: texts/Methodology-GP.tex
\section{The Proposed FPGA Macro Placer}\label{sec:methodology}
Macro placement determines the locations of macro instances, such as DSPs and memories, on the FPGA.
With macro locations fixed, the other non-macro instances are placed and routed to generate a bit stream to program the FPGA.
Thus, the quality of macro placement at the first stage significantly impacts the routability of the entire design.

\begin{figure}[!t]
    %\vspace{-0.1in}
    \centering
    \includegraphics[width=0.65\columnwidth]{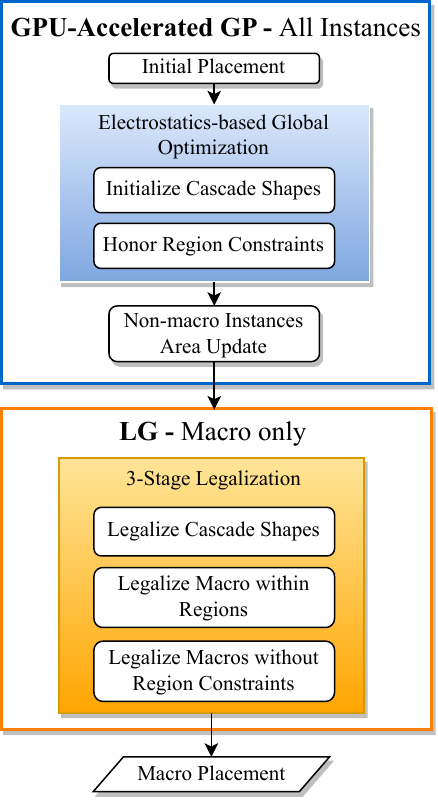} 
    %\vspace{-0.1in}
    \caption{The proposed macro placement flow.}
    \label{fig:Macro_Placement_Flow}
    %\vspace{-0.2in}
\end{figure}

Figure~\ref{fig:Macro_Placement_Flow} presents the two-stage macro placement flow in the proposed macro placer that consists of global placement (GP) and legalization (LG) phases.
The first component is global placement with GPU acceleration, where we employ a mixed-size electrostatics-based global placement approach for all instances.
The second component involves macro legalization. To tackle issues related to cascade shapes and region constraints, we break down this problem into three stages.
Lastly, in the third component, we further refine the legalized solutions and generate the final macro placement solution.

All the design instances are considered to ensure design routability during global placement, and the details are specified in Section~\ref{subsection:GP}.
The global placement stage obtains rough legal locations for all the design instances and operators, such as wirelength and density accelerated on a GPU.
Section~\ref{subsection:MacroLG} describes the three-step legalization of macros to ensure cascade shapes and region constraints are handled correctly.

%

%%%%%%%%%%%%%%%%
\subsection{GPU-accelerated Global Placement} \label{subsection:GP}
The macro placement problem typically involves the placement of a  few thousand macro instances.
However, having an effective placement requires considering the interconnections between all the macros and non-macro instances.
A macro-placement that does not consider non-macro instances could affect the placement of the non-macro instances or result in an unroutable placement.
Including macro and non-macro instances during global placement provides a unified global view of placeable instances, enabling more accurate estimations of the density and congestion profile to ensure a routable design.
\subsubsection{Handling Macro Cascade Shapes}\label{subsubsection:CascadeInit}
A cascade shape consists of several macros placed in a predefined order to ensure proper functionality.
Cascade shapes vary in size, ranging from a few to tens of macro instances, resulting in a cascade shape occupying up to half the column.
As discussed in Section ~\ref{sec:background}, the macro instances in a cascade shape must be consecutively placed in contiguous sites, which resembles a cascaded placement pattern.
In a cascade shape, the $1_{st}$ macro instance drives the $2_{nd}$, and consequently, the outputs from the $2_{nd}$ macro instance drives the $3_{rd}$, creating a sequential chain of connections.

\begin{figure}[!t]
    %\vspace{-0.1in}
    \centering
    \includegraphics[width=0.55\columnwidth]{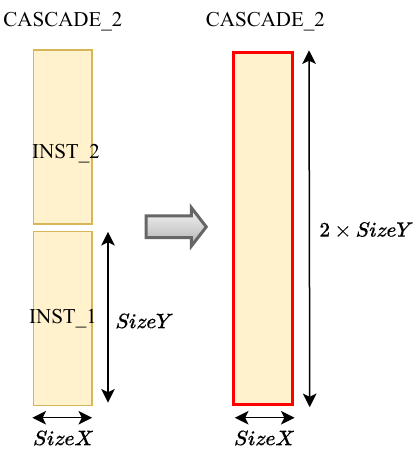}
    %\vspace{-0.2in}
    \caption{Grouping all the macros in a cascade shape as one unified instance.}
    \label{fig:Merge_Cascade}
    \vspace{-0.2in}
\end{figure}

To effectively handle the macro instances in a cascade shape, we propose to first \textit{merge} the individual macro instances into a single cascade unit as also in ~\cite{AMFPlacer, liang2022amf}, then treat the merged entity as a single instance during global placement, as shown in Figure~\ref{fig:Merge_Cascade}.
This unified representation significantly simplifies the placement task for macros in a cascade shape to yield stable results.
Our experiments have consistently shown that merging the macro instances in a cascade shape does not affect global placement convergence and eases the efforts of the legalization stage.

%%%%%%%%%%%%%%%%%%%%%%%%%%%%%%%%
\subsubsection{Handling Region Constraints}\label{subsubsection:boundInsts}
Like ASICs, FPGAs also impose region constraints that restrict specific collections of instances within the designated rectilinear regions on the FPGA, as depicted in Figure ~\ref{fig:Region_Boxes}.
In addition to ensuring instances are placed in corresponding sites, region constraints must be honored.

A straightforward approach to address the region constraints would be to handle the regions separately by incorporating additional electrostatic systems as in ~\cite{DREAMPlace3.0, 10.1145/3489517.3530568}.
Four electrostatic systems are required to handle the LUT, FF, DSP, and BRAM types for instances not subject to any region constraint.
A region constraint spanning $N$ resource types would require $N$ additional electrostatic systems.
For instance, the \textit{Region1} in Figure ~\ref{fig:Region_Boxes} containing instances corresponding to LUT, FF, DSP, and BRAM resource types requires $N=4$ additional electrostatic systems.
A design with three region constraints would require $3 \times 4 = 12$ electrostatic systems and an additional 4 systems to handle instances not assigned to any regions, resulting in 16 electrostatic systems.
As the number of region constraints increases, employing separate electrostatic systems for each region and resource type significantly increases the runtime overhead to define and solve the electrostatic systems for the placement problem.
Therefore, it is necessary to employ a suitable approach that minimizes runtime impact while addressing all the region constraints effectively.

In \sys{}, we ensure that all instances with region constraints align to their respective regions during each iteration of global placement.
This approach applies to any instance (macro and non-macro) with a region constraint.
Consider an instance $i$ of size $w_i \times h_i$ at location $(i_x, i_y)$ that has been assigned to \textit{Region1}: $[R^1_{xl}, R^1_{yl}, R^1_{xh}, R^1_{yh}]$.
The location of instance $i$ is updated during each global placement iteration to ensure it remains in \textit{Region1} as indicated in Equation \eqref{eqn:move_boundary}.
\begin{align}
    \begin{split} \label{eqn:move_boundary}
         &i_x = \max(i_x, R^1_{xl}) \\
        &i_x =\min(i_x, R^1_{xh}-w_i) \\
        &i_y = \max(i_y, R^1_{yl}) \\
        &i_y =\min(i_y, R^1_{yh}-h_i)
    \end{split}
\end{align}

The global placement could diverge when many regions with high utilizations and macro cascade shapes contend for available space on the FPGA, resulting in an infeasible placement.
In such cases, we roll back to a previously saved checkpoint with the best placement to avoid an invalid solution due to divergence during global placement.
The rollback during divergence guarantees a valid routable macro placement solution.

%%%%%%%%%%%%%%%%%%%%%%%%%
\subsubsection{GPU-accelerated Operators} \label{subsubsection:GPU acceleration}
As \sys{} is built on \cite{dreamplacefpga}, several operators, such as wirelength, density, and congestion-aware instance area update for LUTs and FFs, are already accelerated on a GPU.
We implement the region boundary alignment operator specified in Section~\ref{subsubsection:boundInsts} on both CPU and GPU.
GPU acceleration significantly reduces the macro global placement runtime compared to a multi-thread CPU implementation.

%% file: texts/Methodology-LG.tex
\begin{algorithm}[b]
\caption{The 3-Stage Macro Legalization} \label{alg:3stageLG}
\begin{algorithmic}[1]
\Require List of unified cascade shape instances $I_{cas}$, list of macros with region constraints $I_{R}$ with regions $k \in R$, list of remaining macros without any region constrains $I_{rem}$, list of macro site locations $S$, and an initial global placement solution for all macro instances.
\Ensure Legal placements for all macro instances.
\phase{Legalize Cascade Shapes} \label{alg:3stageLG:Cas_greedy_LG}
\State{Sort cascade shape instances from the largest to the smallest.}
\For {each $i \in \text{Sorted } I_{cas}$}                                         
    \State $C \longleftarrow$ list of available candidate sites that can fit $i$ 
    \State $c_{best} \longleftarrow$ closest candidate to $i$ in $C$  
    \State Assign $c_{best}$ as site locations for $i$                                  
    \State Mark $i$ as legalized; Remove $c_{best}$ from $S$                               
\EndFor                                                   
\phase{Legalize Macros within Regions} \label{alg:3stageLG:LG_Regions}
\State {\textbf{Sort regions $R$ based on increasing \#available sites}}  \label{alg:3stageLG:LG_Regions:1}
\For {each $k \in \text{Sorted Regions } R$}
    \State $M_k \longleftarrow$ List of macros in region $I_{R_k}$
    \State $S_k \longleftarrow$ Available sites in region $k$
    \State Solve a bipartite matching for $M_k \longrightarrow S_k$
    \State Mark $I_{R_k}$ as legalized; Remove $S_k$ from $S$
\EndFor                                                      
\phase{Legalize Macros without Region Constraints}  \label{alg:3stageLG:remaining}
\State $M_{rem} \longleftarrow$ Macros without regions $I_{rem}$
\State $S_{rem} \longleftarrow$ Available sites $S$ 
\State Solve a bipartite matching for $M_{rem} \longrightarrow S_{rem}$
\State Mark $I_{rem}$ as legalized
\end{algorithmic}
\end{algorithm}
%
%%%%%%%%%%%%%%%%%%
\subsection{3-Stage Macro Legalization}\label{subsection:MacroLG}
Global placement terminates when the overlap or overflow (OVFL between instances falls below a specified threshold (OVFL < 0.1 for non-macro instances; OVFL < 0.2 for macro instances)~\cite{elfPlace}.
Roughly legal placements are obtained for all the instances at the end of the global placement stage.
The next stage, macro legalization, assigns macro to respective site locations, as depicted in Figure~\ref{fig:Macro_Placement_Flow}.
We employ a three-stage macro legalization as elaborated in Algorithm ~\ref{alg:3stageLG} to ensure legal macro placements with cascade shapes and region constraints.

%%%%%%%%%%%%%%%%%%%%%%
\subsubsection{Legalize Cascade Shapes}
As cascade shapes contain several macro instances and occupy multiple sites, we legalize them first in Phase \ref{alg:3stageLG:Cas_greedy_LG} of macro legalization Algorithm~\ref{alg:3stageLG}.
To ensure all the cascade shapes can be assigned to legal sites, we sort them in decreasing order of their sizes from the largest to the smallest (line 1).
Lines 2-7 describe how each cascade shape is legalized.
For the considered cascade shape $i$, a list of candidate sites $C$ is obtained.
A candidate $c \in C$ consists of multiple sites to accommodate all the macro instances in the cascade shape.
If a cascade shape instance has a region constraint, the candidate sites within the region $C \in R_k$ are selected.
The best candidate $c_{best} \in C$ is closest to the location of $i$ at the end of global placement.
$i$ is legalized to $c_{best}$ and the avaiable macro sites in $S$ are updated.

%%%%%%%%%%%%%%%%%%%%%%%%%
\subsubsection{Legalize Macros within Regions}
After legalizing all the cascade shape instances, we consider macro instances with region constraints in Phase \ref{alg:3stageLG:LG_Regions} of Algorithm~\ref{alg:3stageLG}.
As regions vary in size and can overlap, we sort the regions in ascending order based on the number of available sites in line 8 and legalize macro instances in one region at a time.
This approach guarantees all macros to find legal sites in respective regions irrespective of the size and utilization of a given region.

\begin{figure}[t]
  \centering
  \includegraphics[width=0.85\columnwidth]{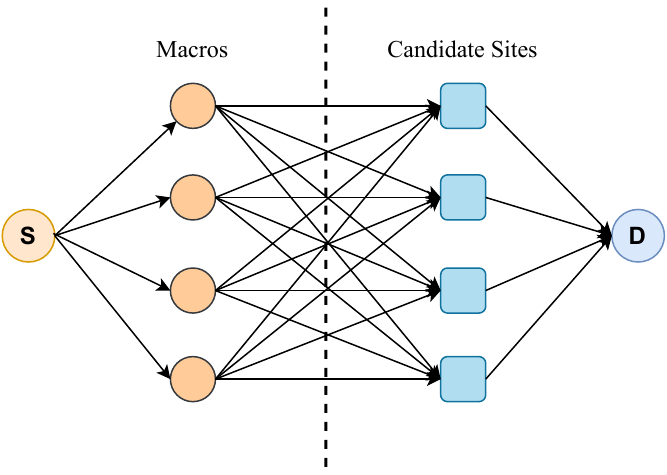}
  \caption{Macros assigned to respective sites using a min-cost flow approach.}
  \Description{}
  \label{fig:Min_cost}
  \vspace{-0.15in}
\end{figure}

The macro instances $M_k$ within a region $R_k$ are assigned to sites $S_k$ using a bipartite matching formulation that is solved using a min-cost flow approach similar to \cite{UTPlaceF, elfPlace} (lines 9-14).
A $M_k \times S_k$ graph is constructed to solve a min-cost flow formulation to assign macro instances to candidate sites, as illustrated in Figure ~\ref{fig:Min_cost}.
The edges connecting macro instances to sites are weighted based on the distance of the macro instance from the site.
An edge cost for a macro instance $m \in M_k$ to a site $s \in S_k$ is computed as follows:
\begin{equation} \label{eqn:Arc_cost}
    \begin{aligned}
        cost(m, s) &= \alpha \cdot precondWL_m \cdot dist(m, s) \\
        precondWL_m &= \left( \sum_{e\in E_m} \frac{1}{|e|-1} \right)
    \end{aligned}
\end{equation}
where $\alpha$ is a scaling factor empirically set to $100$, $dist(m,s)$ is the Manhattan distance between the macro $m$ location at the end of global placement to site $s$, and $precondWL_m$ is the preconditioner.
The preconditioning term $precondWL_m$ depends on the number of connections of all the instance nets $e \in E_m$, and $|e|$ refers to the number of pins connected to net $e$.
An instance with more connections has a larger preconditioner value than an instance with fewer connections.

By solving a min-cost flow problem, the macro instances $M_k$ are assigned to corresponding sites in $S_k$.
The assigned sites in $S_k$ are removed from the available macro sites $S$ list in line 13.
%
% Min-cost flow for LG
%%%%%%%%%%%%%%%%%%%%%%%%
\subsubsection{Legalize Macros without Region Constraints}
After legalizing cascade shape instances and macro instances with region constraints, we legalize all the remaining macro instances that do not have any region constraints in Phase \ref{alg:3stageLG:remaining} of Algorithm~\ref{alg:3stageLG}.
Similar to Phase \ref{alg:3stageLG:LG_Regions}, a bipartite matching is used to assign macro instances $M_{rem}$ to the available sites $S_{rem}$.
A $M_{rem} \times S_{rem}$ graph is constructed with an edge $cost(m,s)$ in Equation \eqref{eqn:Arc_cost}, which is then solved using a min-cost flow approach to assign macro instances to available sites.

After legalizing all the macro instances, a legal macro placement is generated by \sys{}.
With a macro placement solution, the subsequent steps include placing all the non-macro instances, routing the design, and generating the bit-stream to program the FPGA. 

%% file: texts/Experiments.tex
\section{Experimental Results}\label{sec:exp}
This section evaluates the proposed open-source macro placer\footnote{\href{https://github.com/zhilix/DREAMPlaceFPGA-MP}{https://github.com/zhilix/DREAMPlaceFPGA-MP}} using the MLCAD 2023 FPGA Macro-Placement Contest benchmarks ~\cite{mlcad2023}, specifically on the Xilinx Ultrascale+ FPGA part xcvu3p-ffvc1517-1-i.
Subsection ~\ref{subsection:expt_setup} outlines the benchmarks, evaluation metrics and experimental setup. 
Subsection ~\ref{subsection:Overall_pref} presents the overall performance of our macro placer in macro placement solution comparisons with other contest participants.
Subsection ~\ref{subsection:vivado_comp} assesses \sys{}'s final placement HPWL, runtime, and routability compared to Vivado's Place-and-Route (P\&R) Flow.

%%%%%%%%%%%%%%%%%%%%%%%%%
\subsection{Benchmarks \& Setup}\label{subsection:expt_setup}
\begin{figure*}[!t] 
    %\vspace{-0.1in}
    \centering
    \includegraphics[width=1.0\linewidth]{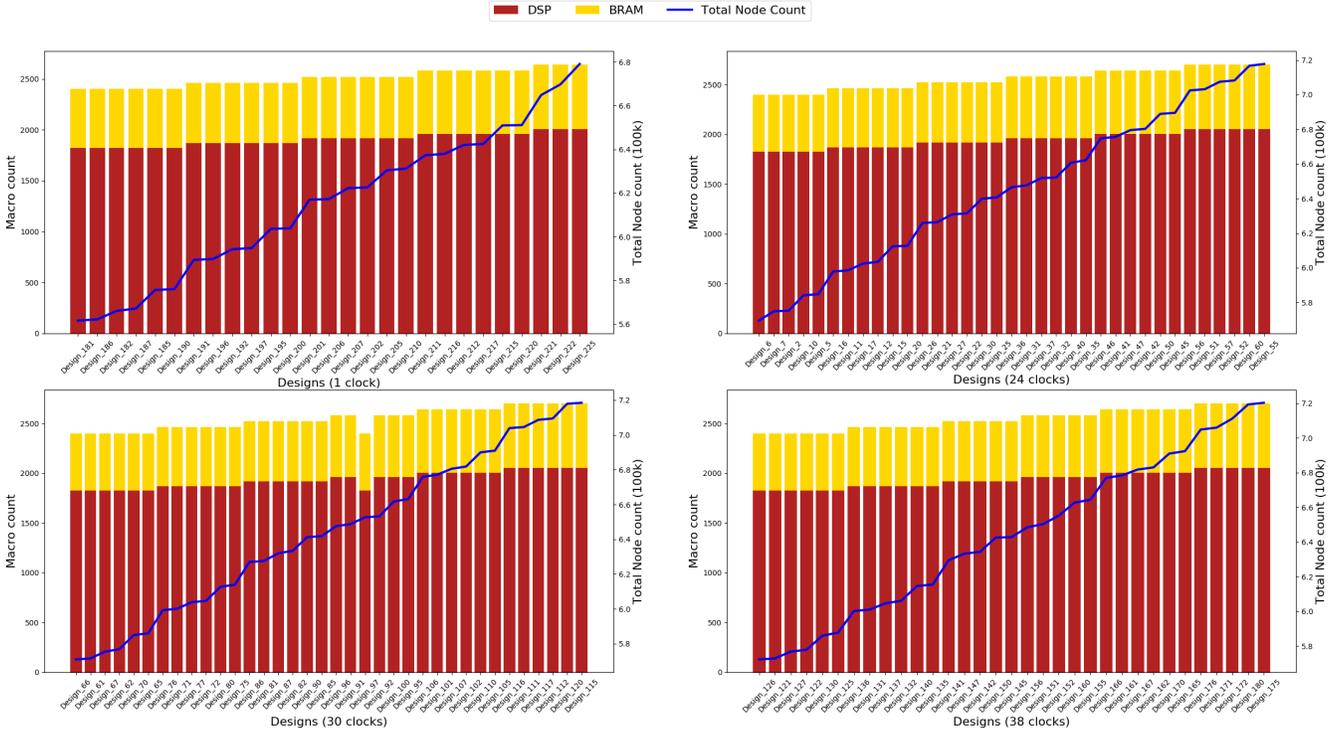}
    \vspace{-0.25in}
    \caption{Stasticis of designs in MLCAD 2023 contest public benchmarks, including number of macros and instances.}
    \label{fig:mlcad_designs}
    \vspace{-0.1in}
\end{figure*}

%%%%%%%%%%%%%%%%%%%%%%%%
\noindent\textbf{Macro Placement Contest}.
The benchmark set for the MLCAD 2023 FPGA Macro Placement Contest~\cite{mlcad2023} consists of 140 public benchmarks of varying size and clocks as shown in Figure \ref{fig:mlcad_designs}.
Additionally, there are 198 hidden benchmarks that have not been disclosed.
Benchmarks encompass designs with 1-clock, 24-clocks, 30-clocks, and 38-clocks. 
In these benchmarks, the total number of instances ranges from 560k to 720k, with an aggregate count of approximately 2,500 macro nodes. 
Within the context of this contest, the netlists exclusively incorporate DSPs and BRAMs as macro elements, deliberately excluding URAMs. 
Additionally, the Cascade shapes featured in this contest exclusively consist of DSPs and BRAMs, with 10 different sizes.

%%%%%%%%%%%%%%%%%%%%%%%%
\noindent\textbf{Evaluation Metrics}.
We consider the runtime and routability scores according to the contest's objectives based on wire length-driven placement that does not consider timing optimization.
The following steps were executed to obtain the evaluation scores:
1) \sys{} was employed to generate legal macro placements;
2) After determining the macro positions, Vivado 2021.1 was used to place non-macro instances;
3) Routing was run on Vivado to determine routability and congestion.
% %
The contest's evaluation score is defined by Equation ~\eqref{eqn:Final_score}, where $i$ is a specific design.
$t'_{MP}$ is the macro placement runtime score, $t_{P\&R}$ is the runtime for place and route in the scale of hours, and $\rho$ is the routability.
Specifically, a lower score indicates better performance.

\begin{equation} \label{eqn:Final_score}
    \begin{aligned}
     Score(i) = t'_{MP}(i) \times t_{P\&R}(i) \times {\rho}(i)
    \end{aligned}
\end{equation}

The macro placement runtime score $t'_{MP}$ is derived from the actual macro placement runtime $t_{MP}$ in the scale of minutes. To ensure timely solutions, a penalty is imposed for macro placement runtimes exceeding 10 minutes, as indicated in Equation ~\eqref{eqn:runtime_macro}.

\begin{equation} \label{eqn:runtime_macro}
    \begin{aligned}
     t'_{MP} =  1 + \max(0, t_{MP}-10)
    \end{aligned}
\end{equation}

The routability score consists of initial routing and detailed routing components, as Equation ~\eqref{eqn:cong_score} describes.
$Sr_i$ represents the initial routing score from global routing by Vivado that measures routing feasibility.
$Sr_f$ is the final routing score during detailed routing that comprehensively evaluates routability and congestion after Vivado's router optimizations.

\begin{equation} \label{eqn:cong_score}
    \begin{aligned}
     \rho = Sr_i + Sr_f 
    \end{aligned}
\end{equation}

As shown in Equation ~\eqref{eqn:init_cong}, the initial routing congestion score captures the short and global congestion levels, denoted as $L^{short}_i$ and $L^{global}_i$, in the $i$-th direction, reported by the initial routing phase in Vivado.
We have four directions: North, South, East, and West, and we sum their score up to get the thorough evaluation metric.

\begin{equation} \label{eqn:init_cong}
    \begin{aligned}
     Sr_i = 1 + \sum_{i=1}^4(\max(0, L^{short}_i-3)^2 + \max(0, L^{global}_i-3)^2)
    \end{aligned}
\end{equation}

The final routing score in Equation ~\eqref{eqn:final_cong} is evaluated by countering the detailed router outer iterations, $\#{dri}$.

\begin{equation} \label{eqn:final_cong}
    \begin{aligned}
     Sr_f = \#{dri}
    \end{aligned}
\end{equation}

%%%%%%%%%%%%%%%%%%%%%%%%%
\noindent\textbf{Detailed Setup}.
In Subsection ~\ref{subsection:Overall_pref}, the runtime scores for all the macro placers and Vivado 2021.1 place and route, were assessed on standalone servers with the following specifications: 3.885GHz CPU, 512GB RAM, and a 64-processor AMD EPYC 7F52 by the contest organizers.
In Subsection ~\ref{subsection:vivado_comp}, we compare the CPU and GPU versions of \sys{} with Vivado 2021.1 on the 140 public benchmarks on an Intel i7-12700 CPU and an NVIDIA TITAN Xp GPU.

\subsection{Comparison with Contest Participants} \label{subsection:Overall_pref}

Based on the runtime and routability scores on the public and hidden benchmarks of the MLCAD 2023 FPGA Macro-Placement Contest, our macro placer was assessed against 5 others: CUMP, CUMPLE, MPKU, SEU, and TAMU, as the contest finalists.
It is important to note that the results from team MPKU's updated submission after the contest are not included in the ranking.
 
In the context of the 140 released public benchmarks and the yet-to-be-released 38 out of 198 hidden benchmarks, our macro placer obtains the lowest/best weighted scores, as outlined in Equations ~\eqref{eqn:Weighted_score} and ~\eqref{eqn:Weights}, and indicated in Table ~\ref{tab:Weighted_scores}.

\begin{equation} \label{eqn:Weighted_score}
    \begin{aligned}
     Final\ Score = \frac{\sum_{i=1}^{\# design}w(i)\cdot Score^2(i)}{\sum_{i=1}^{\# design} w(i)}
    \end{aligned}
\end{equation}

\begin{equation} \label{eqn:Weights}
    \begin{aligned}
     w(i) = \begin{cases} 
     1 & \text{$i \in$ public benchmarks} \\
     140/38 & \text{$i \in$ hidden benchmarks}
     \end{cases}
    \end{aligned}
\end{equation}

\begin{table}[htbp]
    \centering
    \caption{Weighted scores of public and hidden benchmarks}
    \label{tab:Weighted_scores}
    \resizebox{0.4\columnwidth}{!}{
    \begin{tabular}{|c|c|c|}
        \hline
        \textbf{Team} & \textbf{Final Score}\\
        \hline
        \textbf{Ours}& \textbf{2.513} \\
        \hline
        \textbf{MPKU}& 2.751\\
        \hline
        \textbf{SEU}&  2.516\\
        \hline
        \textbf{TAMU}&  4.399\\
        \hline
        \textbf{CUMPLE}&  3.605\\
        \hline
        \textbf{CUMP}&  8.433\\
        \hline
    \end{tabular}
    }
\end{table}

\begin{figure}[h] 
    %\vspace{-0.1in}
    \centering
    \includegraphics[width=\linewidth]{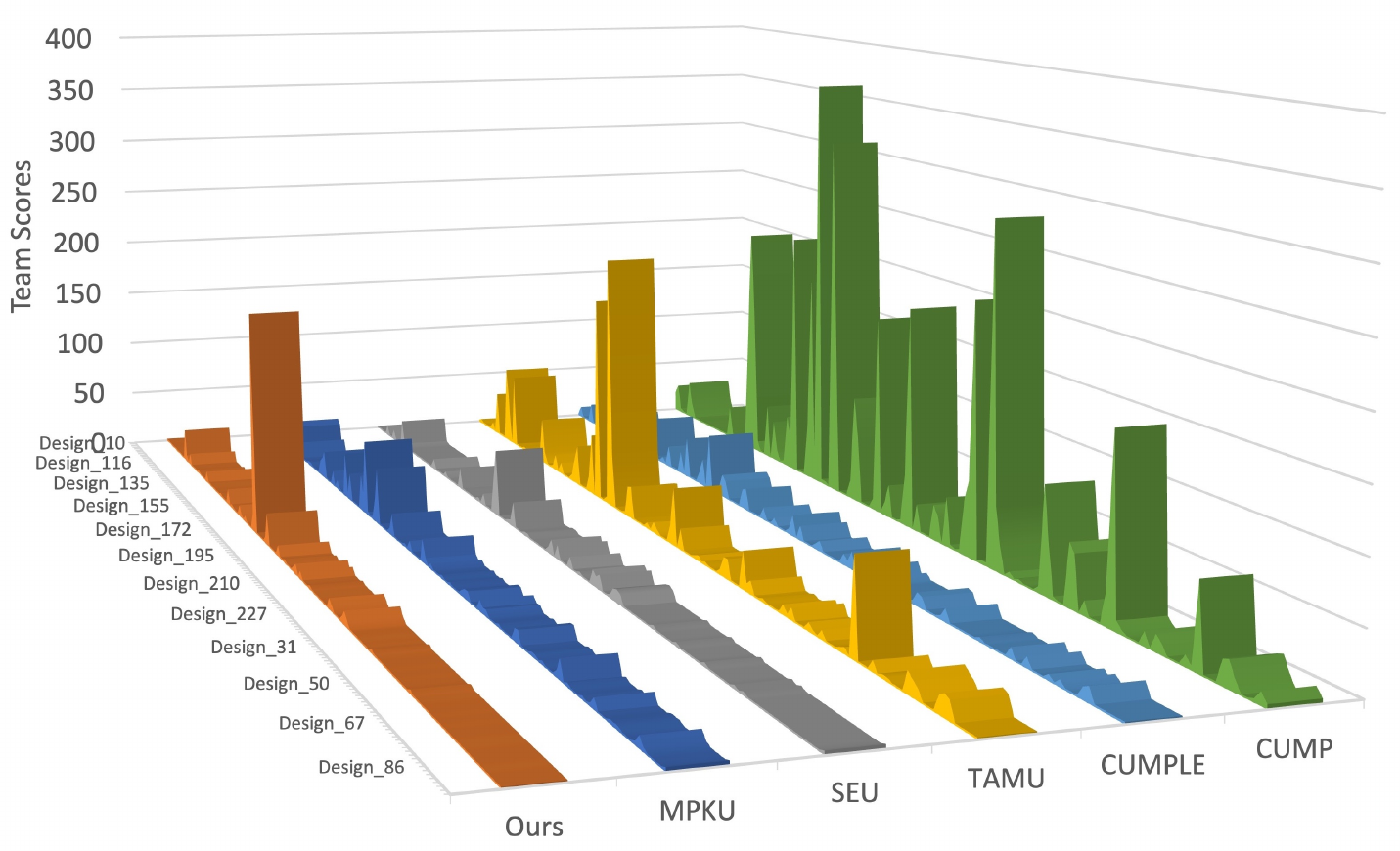}
    \vspace{-0.25in}
    \caption{Overall Scores on the MLCAD 2023 contest public benchmarks.}
    \label{fig:Public_Designs_Scores}
    \vspace{-0.1in}
\end{figure}

The overall scores on the 140 public benchmarks are listed in Table ~\ref{tab:Public_benchmarks} and plotted in Figure ~\ref{fig:Public_Designs_Scores} for all the finalists.
While our macro placer achieves the lowest geomean score across all the 140 public benchmarks, we observe that team SEU achieves the lowest average score and standard deviation (Stddev).
In  Figure ~\ref{fig:Public_Designs_Scores}, we observe that while our placer generally achieves lower (and thus better) scores than other teams, there are few specific designs for which our macro placer achieves very high scores.
To understand the corner case scenarios in our macro placer performance, we consider runtime and routability components of the final scores.
\begin{table}[htbp]
    \centering
    \caption{Score summary on the 140 public designs in MLCAD contest public benchmark}
    \label{tab:Public_benchmarks}
    \resizebox{\columnwidth}{!}{
    \begin{tabular}{|c|c|c|c|c|c|c|}
        \hline
        \textbf{ } & \textbf{Ours} & \textbf{MPKU}  & \textbf{SEU}  & \textbf{TAMU}  & \textbf{CUMPLE} & \textbf{CUMP}\\
        \hline
        Average& 4.204 & 5.716& 3.675& 10.775& 6.188 & 30.603 \\
        \hline
        GeoMean & \textbf{2.120} & 3.023 &2.233 & 3.593 & 3.029 & 6.650\\
        \hline
        Stddev&  16.236& 9.181& 6.495& 26.859& 10.648& 63.609\\
        \hline
    \end{tabular}
    }
\end{table}

\subsubsection{Runtime}
The runtime score comprises two components: the time it takes for a macro placer to generate a valid macro placement solution and the total runtime of Vivado 2021.1 to place all other non-macro instances and route the design.

\begin{figure}[h]
    %\vspace{-0.1in}
    \centering
    \includegraphics[width=\linewidth]{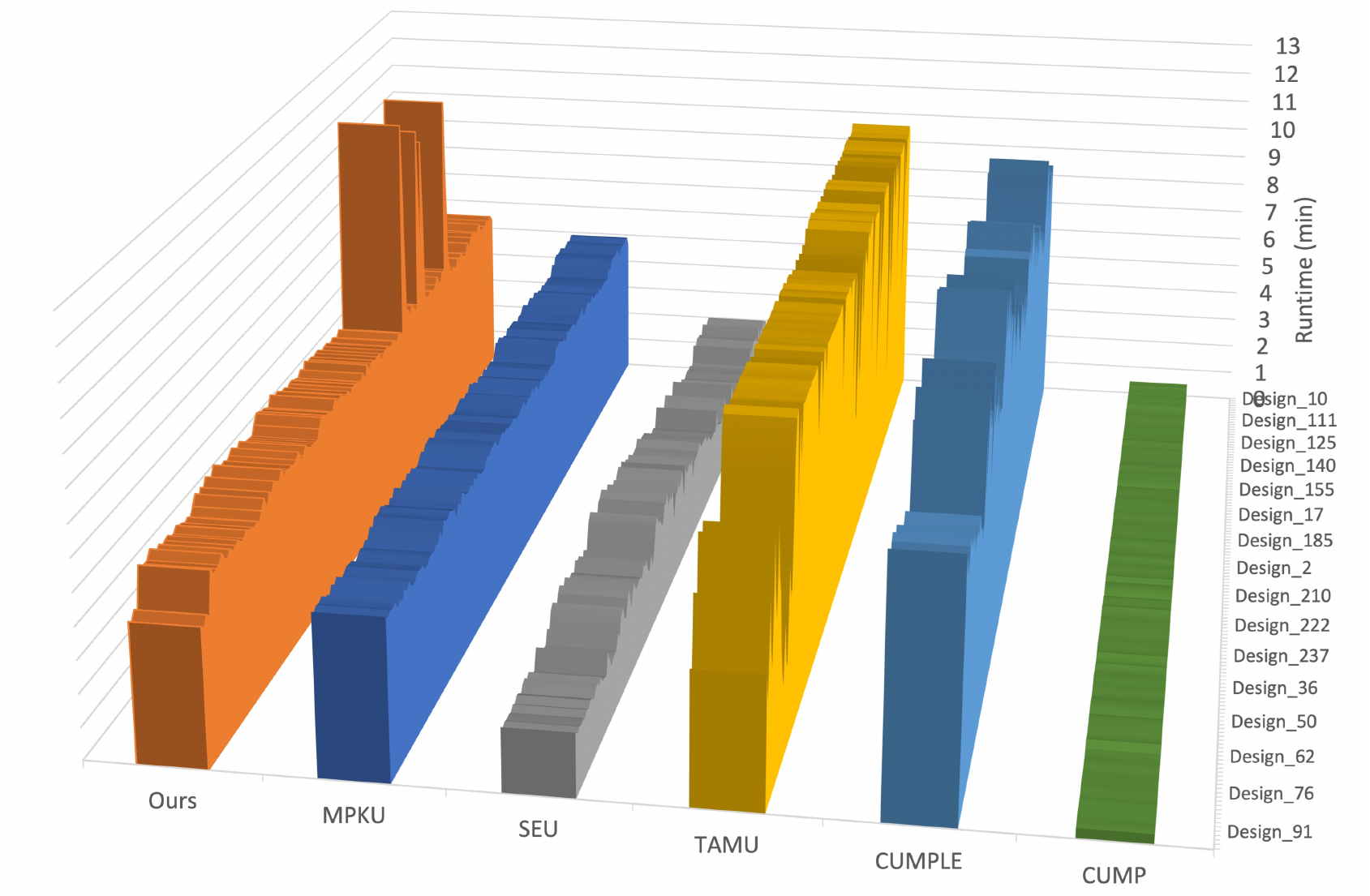}
    %\vspace{-0.1in}
    \caption{Macro placement runtime of MLCAD 2023 contest public benchmarks.}
    \label{fig:Public_Macro_Runtime}
    \vspace{-0.1in}
\end{figure}

Figure ~\ref{fig:Public_Macro_Runtime} compares the macro placement runtime of the CPU version of our macro placer with other teams.
As per Equation~\ref{eqn:runtime_macro}, a runtime penalty is incurred if the macro placement runtime exceeds 10 minutes.
If macro placement completes within 10 minutes, the score $t'_{MP} = 1$.
Most teams achieve a macro placement runtime within the 10-minute target, and Team CUMP has the fastest runtime.
Nevertheless, the contest runtime score does not reward fast macro placement.
For a few designs, as shown in Figure~\ref{tab:Public_benchmarks}, our macro placer exceeds the 10-minute target runtime, which resulted in some penalties in our score.

\begin{figure}[h]
    \centering
    \includegraphics[width=\linewidth]{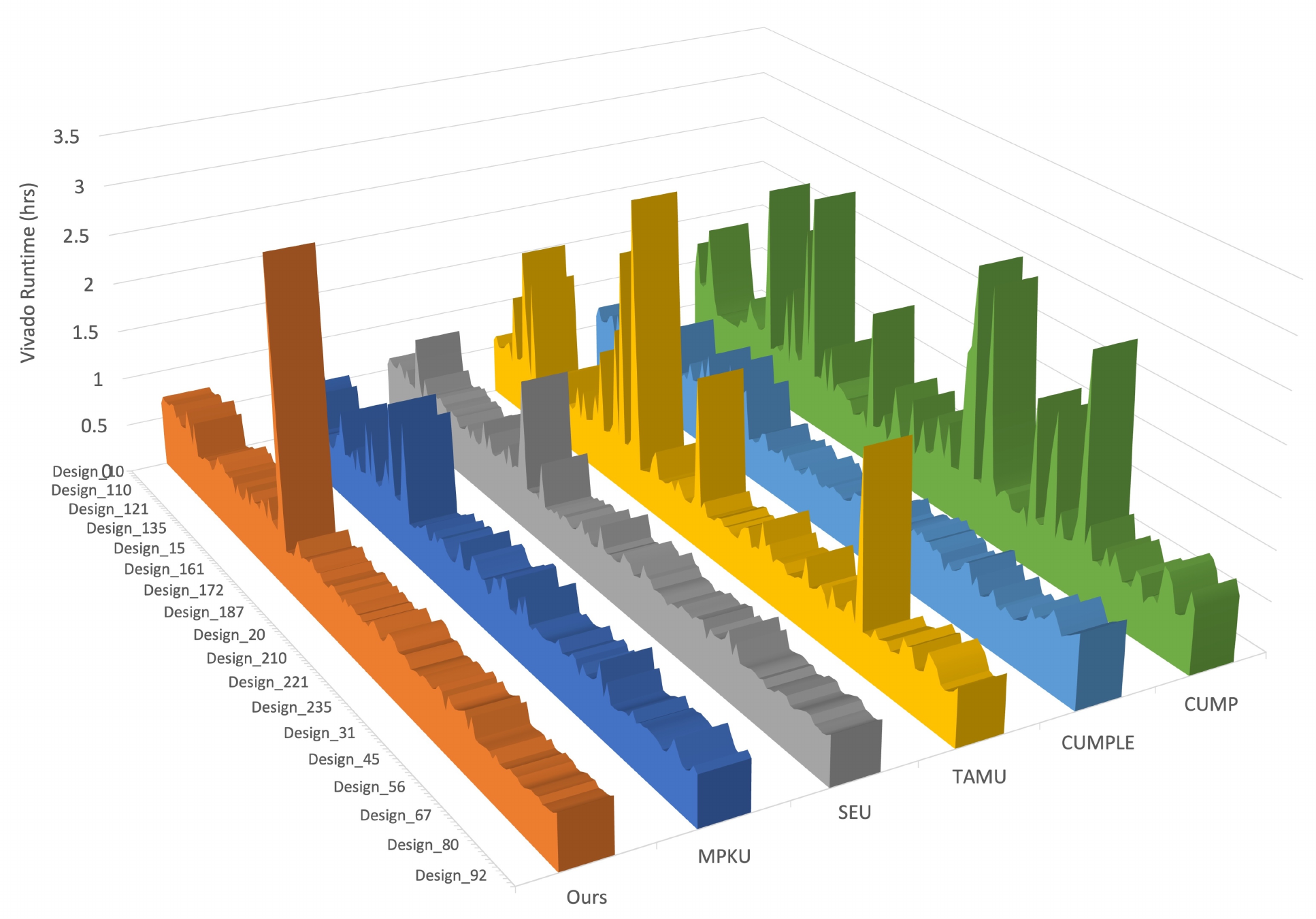}
    % \vspace{-0.1in}
    \caption{Vivado Place-and-Route runtimes on MLCAD 2023 contest public benchmarks.}
    \label{fig:Public_PnR_Runtime}
    \vspace{-0.1in}
\end{figure}

Figure ~\ref{fig:Public_PnR_Runtime} depicts the Vivado 2021.1 Place and Route runtimes (in hours) on the public benchmarks.
It can be noticed that there are remarkably similar peaks in runtime for the same set of designs for all the teams in comparison to the overall scores in Figure~\ref{fig:Public_Designs_Scores}.
Our macro placer has divergence during global placement for 5 out of the 140 public benchmarks, resulting in a large runtime to roll back to a legal macro placement solution as discussed in Section ~\ref{subsubsection:boundInsts}.
Section~\ref{subsubsection:Vivado_comp_runtime} delves into the details of these designs with divergence. 
Our macro placer does not optimize these 5 designs with divergence well, resulting in higher congestion and routability issues when imported into Vivado for the place and route process.
Nevertheless, all the macro placements from our placer for the 140 designs were successfully placed and routed by Vivado.

\begin{figure}[h]
    %\vspace{-0.1in}
    \centering
    \includegraphics[width=\linewidth]{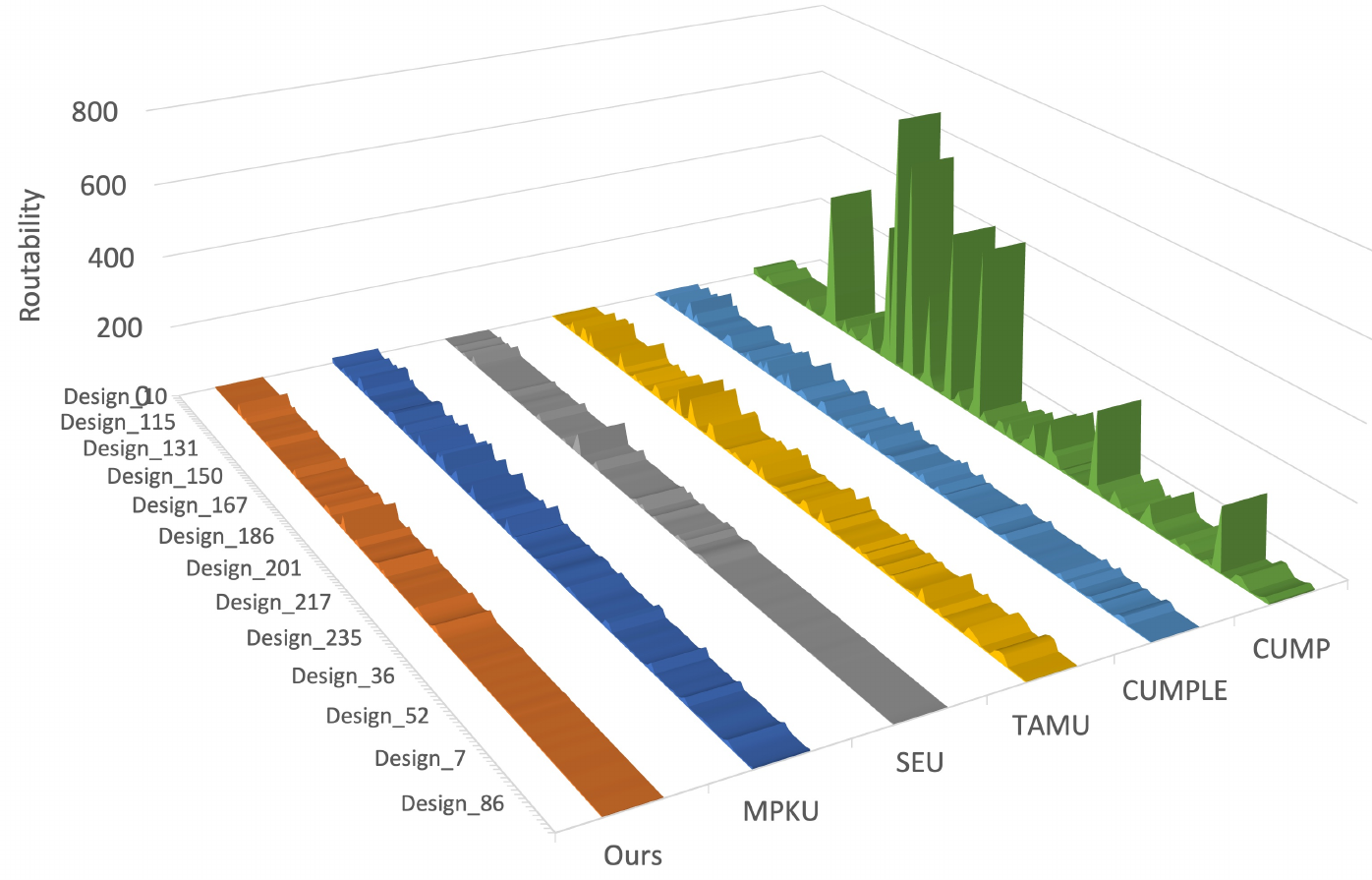}
    \vspace{-0.1in}
    \caption{MLCAD 2023 contest public benchmark routability scores.}
    \label{fig:Routability}
    \vspace{-0.1in}
\end{figure}
\subsubsection{Routability}

The routability score is determined based on the performance of Vivado's router after placement.
The initial routing score is based on an initial congestion estimation by the global router.
During the detailed routing stage, the router undergoes a process of iterative overlap reduction, which constitutes the final routing score.
A higher number of iterations signifies that the router is allocating more effort to address congestion in specific areas of the design.
% The final routing score is determined by Equation ~\eqref{eqn:final_cong}, which quantifies the number of iterations during the detailed routing stage.

Figure ~\ref{fig:Routability} compares the routability scores of all the finalists on the public benchmarks. 
Except for Team CUMP, all other teams' placement solutions remain routable without significant congestion issues.

\subsubsection{MPKU's Updated Submission}
\begin{figure}[h]
    %\vspace{-0.1in}
    \centering
    \includegraphics[width=\linewidth]{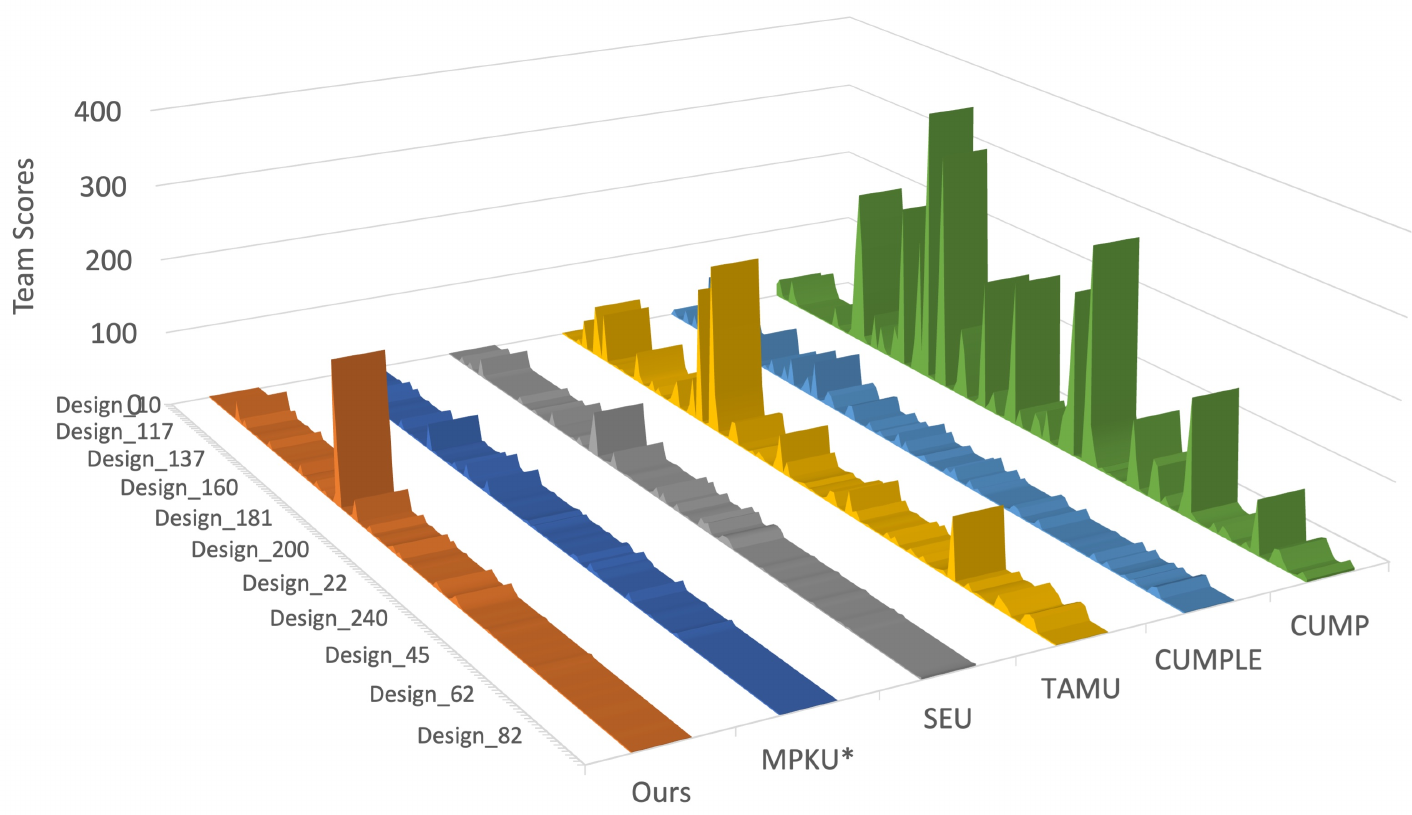}
    \vspace{-0.1in}
    \caption{MLCAD 2023 contest public benchmark team scores with updated submission from MPKU.}
    \label{fig:updated_mpku}
    \vspace{-0.1in}
\end{figure}

After the contest's official deadline, team MPKU submitted an updated version of their macro placer with improvements to runtime and routability.
Figure ~\ref{fig:updated_mpku} compares the overall scores on the 140 public benchmarks with the updated submission from Team MPKU.
Notably, the updated macro placer $MPKU^*$ achieves the best overall score of \textbf{2.216} outperforming all other teams as listed in Table ~\ref{tab:Weighted_scores}.

%%%%%%%%%%%%%%%%%%%%%%%%
\subsection{Comparison With Vivado} \label{subsection:vivado_comp}
%%%%%%%%%%%%%%%%%%%%%%%%%
We compare our macro placer \sys{} with Vivado 2021.1 Place-and-Route (P\&R) flow on the 140 public benchmarks.
In Vivado's default placement flow, macro placement initiates during the partition-based placement (PBP) phase, and the final macro locations are determined in the detailed placement phase.
As mentioned in the detailed setup outlined in Section~\ref{subsection:expt_setup}, we independently carried out the Vivado experiments, and the data may not match the contest evaluation results.
Nevertheless, it offers a fair basis for comparing our macro placer with Vivado.

\begin{table*}[]
\centering
\caption{Comparison of Placement HPWL ($\times 10^3$)}
\label{tab:hpwl_comp}
\vspace{-0.1in}
\resizebox{0.95\textwidth}{!}{%
\begin{tabular}{|cccccccccc|c|c|}
\hline
\multicolumn{1}{|c|}{\textbf{Design}} & \multicolumn{1}{c|}{\textbf{Vivado}} & \multicolumn{1}{c|}{\textbf{MP + V}} & \multicolumn{1}{c|}{\textbf{Design}} & \multicolumn{1}{c|}{\textbf{Vivado}} & \multicolumn{1}{c|}{\textbf{MP + V}} & \multicolumn{1}{c|}{\textbf{Design}} & \multicolumn{1}{c|}{\textbf{Vivado}} & \multicolumn{1}{c|}{\textbf{MP + V}} & \textbf{Design} & \textbf{Vivado} & \textbf{MP + V} \\ \hline
\hline 
\multicolumn{1}{|c|}{Design\_10} & \multicolumn{1}{c|}{7833} & \multicolumn{1}{c|}{6941} & \multicolumn{1}{c|}{Design\_152} & \multicolumn{1}{c|}{6677} & \multicolumn{1}{c|}{6035} & \multicolumn{1}{c|}{Design\_207} & \multicolumn{1}{c|}{7558} & \multicolumn{1}{c|}{7235} & Design\_46 & 8201 & 7847 \\ \hline
\multicolumn{1}{|c|}{Design\_100} & \multicolumn{1}{c|}{8113} & \multicolumn{1}{c|}{7303} & \multicolumn{1}{c|}{Design\_155} & \multicolumn{1}{c|}{6696} & \multicolumn{1}{c|}{5959} & \multicolumn{1}{c|}{Design\_21} & \multicolumn{1}{c|}{5168} & \multicolumn{1}{c|}{4865} & Design\_47 & 7857 & 7701 \\ \hline
\multicolumn{1}{|c|}{Design\_101} & \multicolumn{1}{c|}{6741} & \multicolumn{1}{c|}{6218} & \multicolumn{1}{c|}{Design\_156} & \multicolumn{1}{c|}{9412} & \multicolumn{1}{c|}{8887} & \multicolumn{1}{c|}{Design\_210} & \multicolumn{1}{c|}{6630} & \multicolumn{1}{c|}{6011} & Design\_5 & 5500 & 5069 \\ \hline
\multicolumn{1}{|c|}{Design\_102} & \multicolumn{1}{c|}{6467} & \multicolumn{1}{c|}{5839} & \multicolumn{1}{c|}{Design\_16} & \multicolumn{1}{c|}{7589} & \multicolumn{1}{c|}{7169} & \multicolumn{1}{c|}{Design\_211} & \multicolumn{1}{c|}{5579} & \multicolumn{1}{c|}{5159} & Design\_50 & 8021 & 7624 \\ \hline
\multicolumn{1}{|c|}{Design\_105} & \multicolumn{1}{c|}{6699} & \multicolumn{1}{c|}{6113} & \multicolumn{1}{c|}{Design\_160} & \multicolumn{1}{c|}{9502} & \multicolumn{1}{c|}{8947} & \multicolumn{1}{c|}{Design\_212} & \multicolumn{1}{c|}{5587} & \multicolumn{1}{c|}{5331} & Design\_51 & 5981 & 5608 \\ \hline
\multicolumn{1}{|c|}{Design\_106} & \multicolumn{1}{c|}{8472} & \multicolumn{1}{c|}{7837} & \multicolumn{1}{c|}{Design\_161} & \multicolumn{1}{c|}{6854} & \multicolumn{1}{c|}{6336} & \multicolumn{1}{c|}{Design\_215} & \multicolumn{1}{c|}{6172} & \multicolumn{1}{c|}{5574} & Design\_52 & 5886 & 5895 \\ \hline
\multicolumn{1}{|c|}{Design\_107} & \multicolumn{1}{c|}{8907} & \multicolumn{1}{c|}{7700} & \multicolumn{1}{c|}{Design\_162} & \multicolumn{1}{c|}{7993} & \multicolumn{1}{c|}{7281} & \multicolumn{1}{c|}{Design\_216} & \multicolumn{1}{c|}{7161} & \multicolumn{1}{c|}{6874} & Design\_55 & 6595 & 5826 \\ \hline
\multicolumn{1}{|c|}{Design\_11} & \multicolumn{1}{c|}{5318} & \multicolumn{1}{c|}{5050} & \multicolumn{1}{c|}{Design\_165} & \multicolumn{1}{c|}{7116} & \multicolumn{1}{c|}{6309} & \multicolumn{1}{c|}{Design\_217} & \multicolumn{1}{c|}{7569} & \multicolumn{1}{c|}{7091} & Design\_56 & 7396 & 7015 \\ \hline
\multicolumn{1}{|c|}{Design\_110} & \multicolumn{1}{c|}{9567} & \multicolumn{1}{c|}{8382} & \multicolumn{1}{c|}{Design\_166} & \multicolumn{1}{c|}{8850} & \multicolumn{1}{c|}{8244} & \multicolumn{1}{c|}{Design\_22} & \multicolumn{1}{c|}{5383} & \multicolumn{1}{c|}{5042} & Design\_57 & 9043 & 8479 \\ \hline
\multicolumn{1}{|c|}{Design\_111} & \multicolumn{1}{c|}{6199} & \multicolumn{1}{c|}{5992} & \multicolumn{1}{c|}{Design\_167} & \multicolumn{1}{c|}{9291} & \multicolumn{1}{c|}{8485} & \multicolumn{1}{c|}{Design\_220} & \multicolumn{1}{c|}{6859} & \multicolumn{1}{c|}{6231} & Design\_6 & 6350 & 6228 \\ \hline
\multicolumn{1}{|c|}{Design\_112} & \multicolumn{1}{c|}{6794} & \multicolumn{1}{c|}{6453} & \multicolumn{1}{c|}{Design\_17} & \multicolumn{1}{c|}{7736} & \multicolumn{1}{c|}{7324} & \multicolumn{1}{c|}{Design\_221} & \multicolumn{1}{c|}{5771} & \multicolumn{1}{c|}{5414} & Design\_60 & 8295 & 7529 \\ \hline
\multicolumn{1}{|c|}{Design\_115} & \multicolumn{1}{c|}{6607} & \multicolumn{1}{c|}{6382} & \multicolumn{1}{c|}{Design\_170} & \multicolumn{1}{c|}{10214} & \multicolumn{1}{c|}{9536} & \multicolumn{1}{c|}{Design\_222} & \multicolumn{1}{c|}{4941} & \multicolumn{1}{c|}{4777} & Design\_61 & 5284 & 4992 \\ \hline
\multicolumn{1}{|c|}{Design\_116} & \multicolumn{1}{c|}{9136} & \multicolumn{1}{c|}{9117} & \multicolumn{1}{c|}{Design\_171} & \multicolumn{1}{c|}{7346} & \multicolumn{1}{c|}{6984} & \multicolumn{1}{c|}{Design\_225} & \multicolumn{1}{c|}{5782} & \multicolumn{1}{c|}{5178} & Design\_62 & 5433 & 5635 \\ \hline
\multicolumn{1}{|c|}{Design\_117} & \multicolumn{1}{c|}{8980} & \multicolumn{1}{c|}{8342} & \multicolumn{1}{c|}{Design\_172} & \multicolumn{1}{c|}{7657} & \multicolumn{1}{c|}{7016} & \multicolumn{1}{c|}{Design\_226} & \multicolumn{1}{c|}{7736} & \multicolumn{1}{c|}{7675} & Design\_65 & 6080 & 5232 \\ \hline
\multicolumn{1}{|c|}{Design\_12} & \multicolumn{1}{c|}{5290} & \multicolumn{1}{c|}{5340} & \multicolumn{1}{c|}{Design\_175} & \multicolumn{1}{c|}{7568} & \multicolumn{1}{c|}{6775} & \multicolumn{1}{c|}{Design\_227} & \multicolumn{1}{c|}{8058} & \multicolumn{1}{c|}{7695} & Design\_66 & 6798 & 6564 \\ \hline
\multicolumn{1}{|c|}{Design\_120} & \multicolumn{1}{c|}{9806} & \multicolumn{1}{c|}{8387} & \multicolumn{1}{c|}{Design\_176} & \multicolumn{1}{c|}{10143} & \multicolumn{1}{c|}{9300} & \multicolumn{1}{c|}{Design\_230} & \multicolumn{1}{c|}{7561} & \multicolumn{1}{c|}{6902} & Design\_67 & 7906 & 7405 \\ \hline
\multicolumn{1}{|c|}{Design\_121} & \multicolumn{1}{c|}{5855} & \multicolumn{1}{c|}{5416} & \multicolumn{1}{c|}{Design\_180} & \multicolumn{1}{c|}{9727} & \multicolumn{1}{c|}{9434} & \multicolumn{1}{c|}{Design\_231} & \multicolumn{1}{c|}{5538} & \multicolumn{1}{c|}{5472} & Design\_7 & 6996 & 6645 \\ \hline
\multicolumn{1}{|c|}{Design\_122} & \multicolumn{1}{c|}{5784} & \multicolumn{1}{c|}{5044} & \multicolumn{1}{c|}{Design\_181} & \multicolumn{1}{c|}{4707} & \multicolumn{1}{c|}{4775} & \multicolumn{1}{c|}{Design\_232} & \multicolumn{1}{c|}{6171} & \multicolumn{1}{c|}{5831} & Design\_70 & 7794 & 7352 \\ \hline
\multicolumn{1}{|c|}{Design\_125} & \multicolumn{1}{c|}{6615} & \multicolumn{1}{c|}{5799} & \multicolumn{1}{c|}{Design\_182} & \multicolumn{1}{c|}{4795} & \multicolumn{1}{c|}{4706} & \multicolumn{1}{c|}{Design\_235} & \multicolumn{1}{c|}{6321} & \multicolumn{1}{c|}{5540} & Design\_71 & 5569 & 5268 \\ \hline
\multicolumn{1}{|c|}{Design\_126} & \multicolumn{1}{c|}{7805} & \multicolumn{1}{c|}{7398} & \multicolumn{1}{c|}{Design\_185} & \multicolumn{1}{c|}{5557} & \multicolumn{1}{c|}{5028} & \multicolumn{1}{c|}{Design\_236} & \multicolumn{1}{c|}{7622} & \multicolumn{1}{c|}{7300} & Design\_72 & 6696 & 5820 \\ \hline
\multicolumn{1}{|c|}{Design\_127} & \multicolumn{1}{c|}{7770} & \multicolumn{1}{c|}{7485} & \multicolumn{1}{c|}{Design\_186} & \multicolumn{1}{c|}{5945} & \multicolumn{1}{c|}{6217} & \multicolumn{1}{c|}{Design\_237} & \multicolumn{1}{c|}{7313} & \multicolumn{1}{c|}{6879} & Design\_75 & 5899 & 5330 \\ \hline
\multicolumn{1}{|c|}{Design\_130} & \multicolumn{1}{c|}{8008} & \multicolumn{1}{c|}{7524} & \multicolumn{1}{c|}{Design\_187} & \multicolumn{1}{c|}{7605} & \multicolumn{1}{c|}{6880} & \multicolumn{1}{c|}{Design\_240} & \multicolumn{1}{c|}{7032} & \multicolumn{1}{c|}{6427} & Design\_76 & 8259 & 7249 \\ \hline
\multicolumn{1}{|c|}{Design\_131} & \multicolumn{1}{c|}{5754} & \multicolumn{1}{c|}{5300} & \multicolumn{1}{c|}{Design\_190} & \multicolumn{1}{c|}{8209} & \multicolumn{1}{c|}{7326} & \multicolumn{1}{c|}{Design\_25} & \multicolumn{1}{c|}{6207} & \multicolumn{1}{c|}{5316} & Design\_77 & 7314 & 7377 \\ \hline
\multicolumn{1}{|c|}{Design\_132} & \multicolumn{1}{c|}{5842} & \multicolumn{1}{c|}{5675} & \multicolumn{1}{c|}{Design\_191} & \multicolumn{1}{c|}{4558} & \multicolumn{1}{c|}{4351} & \multicolumn{1}{c|}{Design\_26} & \multicolumn{1}{c|}{7080} & \multicolumn{1}{c|}{6405} & Design\_80 & 8231 & 7443 \\ \hline
\multicolumn{1}{|c|}{Design\_135} & \multicolumn{1}{c|}{6464} & \multicolumn{1}{c|}{5864} & \multicolumn{1}{c|}{Design\_192} & \multicolumn{1}{c|}{5256} & \multicolumn{1}{c|}{5324} & \multicolumn{1}{c|}{Design\_27} & \multicolumn{1}{c|}{7129} & \multicolumn{1}{c|}{6817} & Design\_81 & 5880 & 5400 \\ \hline
\multicolumn{1}{|c|}{Design\_136} & \multicolumn{1}{c|}{8754} & \multicolumn{1}{c|}{7740} & \multicolumn{1}{c|}{Design\_195} & \multicolumn{1}{c|}{5086} & \multicolumn{1}{c|}{4594} & \multicolumn{1}{c|}{Design\_30} & \multicolumn{1}{c|}{7416} & \multicolumn{1}{c|}{6791} & Design\_82 & 6044 & 5786 \\ \hline
\multicolumn{1}{|c|}{Design\_137} & \multicolumn{1}{c|}{9268} & \multicolumn{1}{c|}{8657} & \multicolumn{1}{c|}{Design\_196} & \multicolumn{1}{c|}{6987} & \multicolumn{1}{c|}{6993} & \multicolumn{1}{c|}{Design\_31} & \multicolumn{1}{c|}{5620} & \multicolumn{1}{c|}{5643} & Design\_85 & 6384 & 5788 \\ \hline
\multicolumn{1}{|c|}{Design\_140} & \multicolumn{1}{c|}{8973} & \multicolumn{1}{c|}{8634} & \multicolumn{1}{c|}{Design\_197} & \multicolumn{1}{c|}{7463} & \multicolumn{1}{c|}{7146} & \multicolumn{1}{c|}{Design\_32} & \multicolumn{1}{c|}{5989} & \multicolumn{1}{c|}{5871} & Design\_86 & 8238 & 7539 \\ \hline
\multicolumn{1}{|c|}{Design\_141} & \multicolumn{1}{c|}{7049} & \multicolumn{1}{c|}{6994} & \multicolumn{1}{c|}{Design\_2} & \multicolumn{1}{c|}{5246} & \multicolumn{1}{c|}{5114} & \multicolumn{1}{c|}{Design\_35} & \multicolumn{1}{c|}{6260} & \multicolumn{1}{c|}{5749} & Design\_87 & 8949 & 8334 \\ \hline
\multicolumn{1}{|c|}{Design\_142} & \multicolumn{1}{c|}{6641} & \multicolumn{1}{c|}{6038} & \multicolumn{1}{c|}{Design\_20} & \multicolumn{1}{c|}{7543} & \multicolumn{1}{c|}{6882} & \multicolumn{1}{c|}{Design\_36} & \multicolumn{1}{c|}{7114} & \multicolumn{1}{c|}{6690} & Design\_90 & 8343 & 7337 \\ \hline
\multicolumn{1}{|c|}{Design\_145} & \multicolumn{1}{c|}{7450} & \multicolumn{1}{c|}{6464} & \multicolumn{1}{c|}{Design\_200} & \multicolumn{1}{c|}{7655} & \multicolumn{1}{c|}{7286} & \multicolumn{1}{c|}{Design\_37} & \multicolumn{1}{c|}{7072} & \multicolumn{1}{c|}{7506} & Design\_91 & 7150 & 6134 \\ \hline
\multicolumn{1}{|c|}{Design\_147} & \multicolumn{1}{c|}{8329} & \multicolumn{1}{c|}{7941} & \multicolumn{1}{c|}{Design\_201} & \multicolumn{1}{c|}{4812} & \multicolumn{1}{c|}{4708} & \multicolumn{1}{c|}{Design\_40} & \multicolumn{1}{c|}{8163} & \multicolumn{1}{c|}{7602} & Design\_92 & 6230 & 6114 \\ \hline
\multicolumn{1}{|c|}{Design\_15} & \multicolumn{1}{c|}{5434} & \multicolumn{1}{c|}{5064} & \multicolumn{1}{c|}{Design\_202} & \multicolumn{1}{c|}{5305} & \multicolumn{1}{c|}{5166} & \multicolumn{1}{c|}{Design\_41} & \multicolumn{1}{c|}{5326} & \multicolumn{1}{c|}{5844} & Design\_95 & 6686 & 6008 \\ \hline
\multicolumn{1}{|c|}{Design\_150} & \multicolumn{1}{c|}{8829} & \multicolumn{1}{c|}{8496} & \multicolumn{1}{c|}{Design\_205} & \multicolumn{1}{c|}{5447} & \multicolumn{1}{c|}{4970} & \multicolumn{1}{c|}{Design\_42} & \multicolumn{1}{c|}{5825} & \multicolumn{1}{c|}{5935} & Design\_96 & 7945 & 7378 \\ \hline
\multicolumn{1}{|c|}{Design\_151} & \multicolumn{1}{c|}{7065} & \multicolumn{1}{c|}{6364} & \multicolumn{1}{c|}{Design\_206} & \multicolumn{1}{c|}{6804} & \multicolumn{1}{c|}{6973} & \multicolumn{1}{c|}{Design\_45} & \multicolumn{1}{c|}{5848} & \multicolumn{1}{c|}{5417} & Design\_97 & 8270 & 8155 \\ \hline
\hline 
\multicolumn{10}{|c|}{\textbf{Overall Geomean Ratio}} & 1.0000 & \textbf{0.9368} \\ \hline
\end{tabular}%
}
\end{table*}

\subsubsection{HPWL}
The final routed wirelength serves as a crucial metric for evaluating the effectiveness of a macro placement solution, particularly in the absence of timing-driven optimization. 
While this metric is not officially introduced in the contest's final evaluation, we use the final placement HPWL as an estimate of routed wirelength due to the unavailability of this information from the Vivado router.

We extract the final placement HPWL as follows: 
After completing the placement of both macro and non-macro instances, we use Vivado's Tcl scripting to obtain the cell locations. 
Subsequently, we load this placement solution into our tool to compute the HPWL.
Table ~\ref{tab:hpwl_comp} illustrates two scenarios: the baseline Vivado Placement (Vivado), where Vivado places both macro and non-macro instances, and the alternative scenario (MP+V), where macro placements are from our tool, and non-macro instances are placed by Vivado. 
According to the results in Table ~\ref{tab:hpwl_comp}, our macro placement, followed by finalization, achieves a $6.32\%$ reduction in placement HPWL compared to the full Vivado placement.

\subsubsection{Runtime}\label{subsubsection:Vivado_comp_runtime}

\begin{figure}[h] 
    \vspace{-0.1in}
    \centering
    \includegraphics[width=\columnwidth]{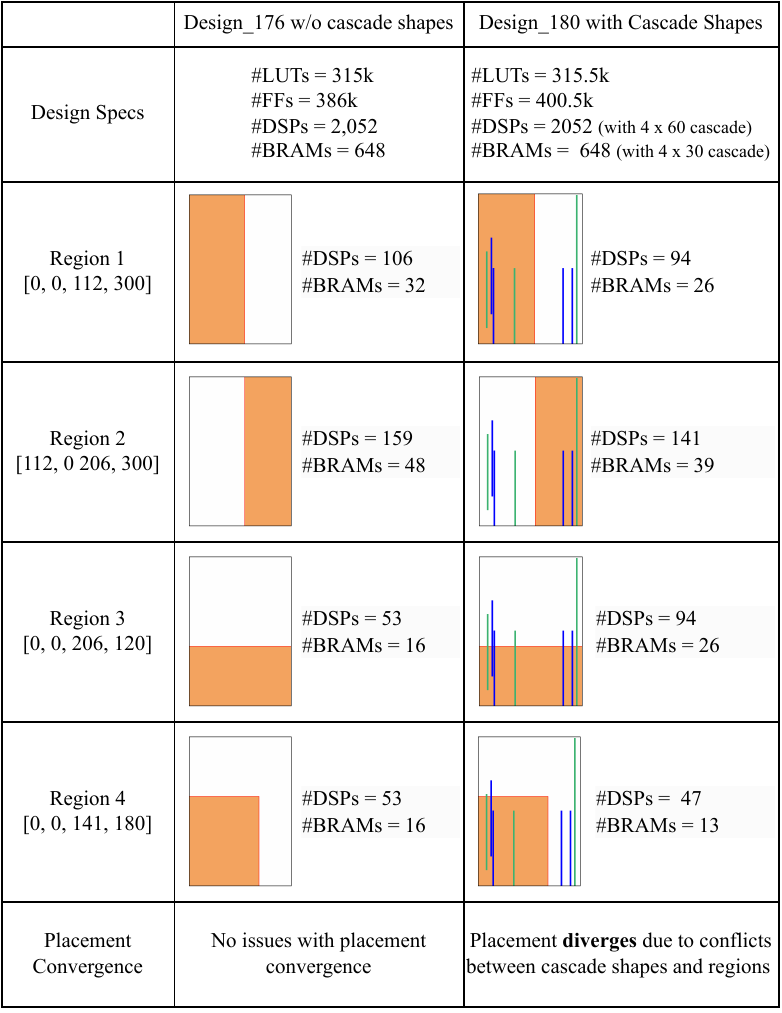}
    \vspace{-0.2in}
    \caption{Impact of Macro Cascade Shapes and Region Constraints on two designs of similar size.}%: Design\_176 vs Design\_180}
    \label{fig:Divergence}
\end{figure}

Similar to the runtime score introduced by the contest, we compare the total placement runtime against Vivado using the proposed macro placer to assess speed and compatibility.
We only consider placement runtime of both \sys{} and Vivado and exclude the data movement runtime.
%We intentionally exclude the IO time from our tool to ensure a fair comparison.
%In the case of baseline Vivado, it only takes the .dcp file once, whereas, for our tool, we need to initiate the design both in Vivado and our tool.
The results for the 140 public benchmarks are presented in Table ~\ref{tab:pl_runtime} and includes GPU and CPU versions of \sys{} for comparison. 
Here, $V_{t_P}$ represents the placement time in Vivado, $t_{MP}$ indicates our macro placement runtime, and $t_{NMP}$ denotes the runtimes from Vivado placement for non-macro instances.
Consequently, we compare $V_{t_P}$ with $Tot$ that is the sum of $t_{MP}$ and $t_{NMP}$ for evaluation.

In summary, across all these designs, the overall placement runtime consisting of macro placer and Vivado placer runtimes is $11.1\%$ and $19.6\%$ faster for CPU and GPU versions of \sys{}, compared to baseline Vivado.

We observe a large total placement runtime using our macro placer for $5$ designs: Design\_140, Design\_141, Design\_162, Design\_166, and Design\_180. 
All these $5$ designs diverge during global placement and roll back to generate a valid placement solution that is not highly optimized.
To understand the reason for the divergence in these designs, we consider a diverging Design\_180 and compare it with another similar design with no divergence issues - Design\_176.
Figure ~\ref{fig:Divergence} compares Design\_176 and Design\_180 designs, which have a similar number of LUTs and FFs and the same number of macros.
While both the designs have 19 region constraints, Design\_180 consists of 4 DSP cascade shapes and 4 BRAM cascade shapes, and Design\_176 contains no cascade macros.
Figure ~\ref{fig:Divergence} illustrates locations of DSP (in \textcolor{blue}{blue}) and BRAM (\textcolor{teal}{green}) cascade shapes at global placement iteration 900, along with 4 regions (in \textcolor{orange}{orange}) containing a large number of macros.
Due to the substantial space occupied by these cascade shape instances in Design\_180, the macros constrained to these regions face challenges in spreading, resulting in increased overlap and divergence in \sys{}.

\renewcommand{\arraystretch}{0.835}% Tighter
\begin{table*}[]
\centering
\vspace{-0.3in}
\caption{Comparison of Placement Runtime ($s$)}
\label{tab:pl_runtime}
\vspace{-0.1in}
\resizebox{0.85\textwidth}{!}{%
\begin{tabular}{|ccccccccc|c|ccc|ccc|}
\hline
\multicolumn{1}{|c|}{\multirow{2}{*}{\textbf{Design}}} & \multicolumn{1}{c|}{\multirow{2}{*}{\textbf{$Vt_P$}}} & \multicolumn{3}{c|}{\textbf{MP (CPU) + V}} & \multicolumn{3}{c|}{\textbf{MP (GPU) + V}} & \multirow{2}{*}{\textbf{Design}} & \multirow{2}{*}{\textbf{$Vt_P$}} & \multicolumn{3}{c|}{\textbf{MP (CPU) + V}} & \multicolumn{3}{c|}{\textbf{MP (GPU) + V}} \\ \cline{3-8} \cline{11-16} 
\multicolumn{1}{|c|}{} & \multicolumn{1}{c|}{} & \multicolumn{1}{c|}{\textbf{$t_{MP}$}} & \multicolumn{1}{c|}{\textbf{$t_{NMP}$}} & \multicolumn{1}{c|}{\textbf{$Tot$}} & \multicolumn{1}{c|}{\textbf{$t_{MP}$}} & \multicolumn{1}{c|}{\textbf{$t_{NMP}$}} & \multicolumn{1}{c|}{\textbf{$Tot$}} &  &  & \multicolumn{1}{c|}{\textbf{$t_{MP}$}} & \multicolumn{1}{c|}{\textbf{$t_{NMP}$}} & \textbf{$Tot$} & \multicolumn{1}{c|}{\textbf{$t_{MP}$}} & \multicolumn{1}{c|}{\textbf{$t_{NMP}$}} & \textbf{$Tot$} \\ \hline
\hline
\multicolumn{1}{|c|}{Design\_10} & \multicolumn{1}{c|}{493} & \multicolumn{1}{c|}{86} & \multicolumn{1}{c|}{325} & \multicolumn{1}{c|}{411} & \multicolumn{1}{c|}{45} & \multicolumn{1}{c|}{343} & \multicolumn{1}{c|}{388} & Design\_207 & 516 & \multicolumn{1}{c|}{94} & \multicolumn{1}{c|}{340} & 434 & \multicolumn{1}{c|}{36} & \multicolumn{1}{c|}{336} & 372 \\ \hline
\multicolumn{1}{|c|}{Design\_100} & \multicolumn{1}{c|}{534} & \multicolumn{1}{c|}{97} & \multicolumn{1}{c|}{412} & \multicolumn{1}{c|}{509} & \multicolumn{1}{c|}{48} & \multicolumn{1}{c|}{412} & \multicolumn{1}{c|}{460} & Design\_21 & 488 & \multicolumn{1}{c|}{85} & \multicolumn{1}{c|}{350} & 435 & \multicolumn{1}{c|}{34} & \multicolumn{1}{c|}{361} & 395 \\ \hline
\multicolumn{1}{|c|}{Design\_101} & \multicolumn{1}{c|}{524} & \multicolumn{1}{c|}{93} & \multicolumn{1}{c|}{372} & \multicolumn{1}{c|}{465} & \multicolumn{1}{c|}{38} & \multicolumn{1}{c|}{402} & \multicolumn{1}{c|}{440} & Design\_210 & 543 & \multicolumn{1}{c|}{95} & \multicolumn{1}{c|}{362} & 457 & \multicolumn{1}{c|}{45} & \multicolumn{1}{c|}{379} & 424 \\ \hline
\multicolumn{1}{|c|}{Design\_102} & \multicolumn{1}{c|}{534} & \multicolumn{1}{c|}{95} & \multicolumn{1}{c|}{375} & \multicolumn{1}{c|}{470} & \multicolumn{1}{c|}{40} & \multicolumn{1}{c|}{391} & \multicolumn{1}{c|}{431} & Design\_211 & 500 & \multicolumn{1}{c|}{94} & \multicolumn{1}{c|}{353} & 447 & \multicolumn{1}{c|}{35} & \multicolumn{1}{c|}{335} & 370 \\ \hline
\multicolumn{1}{|c|}{Design\_105} & \multicolumn{1}{c|}{531} & \multicolumn{1}{c|}{93} & \multicolumn{1}{c|}{383} & \multicolumn{1}{c|}{476} & \multicolumn{1}{c|}{48} & \multicolumn{1}{c|}{384} & \multicolumn{1}{c|}{432} & Design\_212 & 522 & \multicolumn{1}{c|}{93} & \multicolumn{1}{c|}{374} & 467 & \multicolumn{1}{c|}{37} & \multicolumn{1}{c|}{344} & 381 \\ \hline
\multicolumn{1}{|c|}{Design\_106} & \multicolumn{1}{c|}{602} & \multicolumn{1}{c|}{94} & \multicolumn{1}{c|}{413} & \multicolumn{1}{c|}{507} & \multicolumn{1}{c|}{39} & \multicolumn{1}{c|}{411} & \multicolumn{1}{c|}{450} & Design\_215 & 513 & \multicolumn{1}{c|}{94} & \multicolumn{1}{c|}{339} & 433 & \multicolumn{1}{c|}{45} & \multicolumn{1}{c|}{337} & 382 \\ \hline
\multicolumn{1}{|c|}{Design\_107} & \multicolumn{1}{c|}{593} & \multicolumn{1}{c|}{91} & \multicolumn{1}{c|}{418} & \multicolumn{1}{c|}{509} & \multicolumn{1}{c|}{40} & \multicolumn{1}{c|}{420} & \multicolumn{1}{c|}{460} & Design\_216 & 513 & \multicolumn{1}{c|}{94} & \multicolumn{1}{c|}{366} & 460 & \multicolumn{1}{c|}{36} & \multicolumn{1}{c|}{329} & 365 \\ \hline
\multicolumn{1}{|c|}{Design\_11} & \multicolumn{1}{c|}{511} & \multicolumn{1}{c|}{92} & \multicolumn{1}{c|}{336} & \multicolumn{1}{c|}{428} & \multicolumn{1}{c|}{35} & \multicolumn{1}{c|}{334} & \multicolumn{1}{c|}{369} & Design\_217 & 533 & \multicolumn{1}{c|}{96} & \multicolumn{1}{c|}{363} & 459 & \multicolumn{1}{c|}{37} & \multicolumn{1}{c|}{367} & 404 \\ \hline
\multicolumn{1}{|c|}{Design\_110} & \multicolumn{1}{c|}{627} & \multicolumn{1}{c|}{100} & \multicolumn{1}{c|}{439} & \multicolumn{1}{c|}{539} & \multicolumn{1}{c|}{50} & \multicolumn{1}{c|}{437} & \multicolumn{1}{c|}{487} & Design\_22 & 483 & \multicolumn{1}{c|}{86} & \multicolumn{1}{c|}{365} & 451 & \multicolumn{1}{c|}{37} & \multicolumn{1}{c|}{329} & 366 \\ \hline
\multicolumn{1}{|c|}{Design\_111} & \multicolumn{1}{c|}{531} & \multicolumn{1}{c|}{97} & \multicolumn{1}{c|}{392} & \multicolumn{1}{c|}{489} & \multicolumn{1}{c|}{39} & \multicolumn{1}{c|}{397} & \multicolumn{1}{c|}{436} & Design\_220 & 559 & \multicolumn{1}{c|}{96} & \multicolumn{1}{c|}{362} & 458 & \multicolumn{1}{c|}{45} & \multicolumn{1}{c|}{356} & 401 \\ \hline
\multicolumn{1}{|c|}{Design\_112} & \multicolumn{1}{c|}{545} & \multicolumn{1}{c|}{94} & \multicolumn{1}{c|}{393} & \multicolumn{1}{c|}{487} & \multicolumn{1}{c|}{40} & \multicolumn{1}{c|}{399} & \multicolumn{1}{c|}{439} & Design\_221 & 493 & \multicolumn{1}{c|}{94} & \multicolumn{1}{c|}{352} & 446 & \multicolumn{1}{c|}{37} & \multicolumn{1}{c|}{345} & 382 \\ \hline
\multicolumn{1}{|c|}{Design\_115} & \multicolumn{1}{c|}{539} & \multicolumn{1}{c|}{98} & \multicolumn{1}{c|}{406} & \multicolumn{1}{c|}{504} & \multicolumn{1}{c|}{49} & \multicolumn{1}{c|}{406} & \multicolumn{1}{c|}{455} & Design\_222 & 488 & \multicolumn{1}{c|}{96} & \multicolumn{1}{c|}{348} & 444 & \multicolumn{1}{c|}{38} & \multicolumn{1}{c|}{343} & 381 \\ \hline
\multicolumn{1}{|c|}{Design\_116} & \multicolumn{1}{c|}{629} & \multicolumn{1}{c|}{98} & \multicolumn{1}{c|}{468} & \multicolumn{1}{c|}{566} & \multicolumn{1}{c|}{39} & \multicolumn{1}{c|}{455} & \multicolumn{1}{c|}{494} & Design\_225 & 507 & \multicolumn{1}{c|}{102} & \multicolumn{1}{c|}{344} & 446 & \multicolumn{1}{c|}{46} & \multicolumn{1}{c|}{342} & 388 \\ \hline
\multicolumn{1}{|c|}{Design\_117} & \multicolumn{1}{c|}{629} & \multicolumn{1}{c|}{99} & \multicolumn{1}{c|}{444} & \multicolumn{1}{c|}{543} & \multicolumn{1}{c|}{41} & \multicolumn{1}{c|}{450} & \multicolumn{1}{c|}{491} & Design\_226 & 554 & \multicolumn{1}{c|}{97} & \multicolumn{1}{c|}{390} & 487 & \multicolumn{1}{c|}{37} & \multicolumn{1}{c|}{389} & 426 \\ \hline
\multicolumn{1}{|c|}{Design\_12} & \multicolumn{1}{c|}{485} & \multicolumn{1}{c|}{87} & \multicolumn{1}{c|}{340} & \multicolumn{1}{c|}{427} & \multicolumn{1}{c|}{37} & \multicolumn{1}{c|}{331} & \multicolumn{1}{c|}{368} & Design\_227 & 631 & \multicolumn{1}{c|}{100} & \multicolumn{1}{c|}{378} & 478 & \multicolumn{1}{c|}{39} & \multicolumn{1}{c|}{384} & 423 \\ \hline
\multicolumn{1}{|c|}{Design\_120} & \multicolumn{1}{c|}{636} & \multicolumn{1}{c|}{97} & \multicolumn{1}{c|}{455} & \multicolumn{1}{c|}{552} & \multicolumn{1}{c|}{49} & \multicolumn{1}{c|}{453} & \multicolumn{1}{c|}{502} & Design\_230 & 569 & \multicolumn{1}{c|}{101} & \multicolumn{1}{c|}{368} & 469 & \multicolumn{1}{c|}{47} & \multicolumn{1}{c|}{371} & 418 \\ \hline
\multicolumn{1}{|c|}{Design\_121} & \multicolumn{1}{c|}{437} & \multicolumn{1}{c|}{86} & \multicolumn{1}{c|}{321} & \multicolumn{1}{c|}{407} & \multicolumn{1}{c|}{36} & \multicolumn{1}{c|}{321} & \multicolumn{1}{c|}{357} & Design\_231 & 497 & \multicolumn{1}{c|}{96} & \multicolumn{1}{c|}{356} & 452 & \multicolumn{1}{c|}{37} & \multicolumn{1}{c|}{355} & 392 \\ \hline
\multicolumn{1}{|c|}{Design\_122} & \multicolumn{1}{c|}{463} & \multicolumn{1}{c|}{88} & \multicolumn{1}{c|}{319} & \multicolumn{1}{c|}{407} & \multicolumn{1}{c|}{37} & \multicolumn{1}{c|}{320} & \multicolumn{1}{c|}{357} & Design\_232 & 533 & \multicolumn{1}{c|}{101} & \multicolumn{1}{c|}{365} & 466 & \multicolumn{1}{c|}{38} & \multicolumn{1}{c|}{364} & 402 \\ \hline
\multicolumn{1}{|c|}{Design\_125} & \multicolumn{1}{c|}{462} & \multicolumn{1}{c|}{87} & \multicolumn{1}{c|}{326} & \multicolumn{1}{c|}{413} & \multicolumn{1}{c|}{45} & \multicolumn{1}{c|}{349} & \multicolumn{1}{c|}{394} & Design\_235 & 536 & \multicolumn{1}{c|}{101} & \multicolumn{1}{c|}{373} & 474 & \multicolumn{1}{c|}{46} & \multicolumn{1}{c|}{371} & 417 \\ \hline
\multicolumn{1}{|c|}{Design\_126} & \multicolumn{1}{c|}{490} & \multicolumn{1}{c|}{90} & \multicolumn{1}{c|}{369} & \multicolumn{1}{c|}{459} & \multicolumn{1}{c|}{67} & \multicolumn{1}{c|}{375} & \multicolumn{1}{c|}{442} & Design\_236 & 572 & \multicolumn{1}{c|}{98} & \multicolumn{1}{c|}{389} & 487 & \multicolumn{1}{c|}{38} & \multicolumn{1}{c|}{388} & 426 \\ \hline
\multicolumn{1}{|c|}{Design\_127} & \multicolumn{1}{c|}{500} & \multicolumn{1}{c|}{92} & \multicolumn{1}{c|}{354} & \multicolumn{1}{c|}{446} & \multicolumn{1}{c|}{37} & \multicolumn{1}{c|}{371} & \multicolumn{1}{c|}{408} & Design\_237 & 584 & \multicolumn{1}{c|}{100} & \multicolumn{1}{c|}{386} & 486 & \multicolumn{1}{c|}{39} & \multicolumn{1}{c|}{387} & 426 \\ \hline
\multicolumn{1}{|c|}{Design\_130} & \multicolumn{1}{c|}{505} & \multicolumn{1}{c|}{93} & \multicolumn{1}{c|}{362} & \multicolumn{1}{c|}{455} & \multicolumn{1}{c|}{46} & \multicolumn{1}{c|}{367} & \multicolumn{1}{c|}{413} & Design\_240 & 566 & \multicolumn{1}{c|}{99} & \multicolumn{1}{c|}{397} & 496 & \multicolumn{1}{c|}{47} & \multicolumn{1}{c|}{387} & 434 \\ \hline
\multicolumn{1}{|c|}{Design\_131} & \multicolumn{1}{c|}{460} & \multicolumn{1}{c|}{84} & \multicolumn{1}{c|}{331} & \multicolumn{1}{c|}{415} & \multicolumn{1}{c|}{36} & \multicolumn{1}{c|}{336} & \multicolumn{1}{c|}{372} & Design\_25 & 499 & \multicolumn{1}{c|}{90} & \multicolumn{1}{c|}{360} & 450 & \multicolumn{1}{c|}{45} & \multicolumn{1}{c|}{349} & 394 \\ \hline
\multicolumn{1}{|c|}{Design\_132} & \multicolumn{1}{c|}{461} & \multicolumn{1}{c|}{88} & \multicolumn{1}{c|}{332} & \multicolumn{1}{c|}{420} & \multicolumn{1}{c|}{37} & \multicolumn{1}{c|}{336} & \multicolumn{1}{c|}{373} & Design\_26 & 510 & \multicolumn{1}{c|}{85} & \multicolumn{1}{c|}{344} & 429 & \multicolumn{1}{c|}{35} & \multicolumn{1}{c|}{349} & 384 \\ \hline
\multicolumn{1}{|c|}{Design\_135} & \multicolumn{1}{c|}{483} & \multicolumn{1}{c|}{89} & \multicolumn{1}{c|}{341} & \multicolumn{1}{c|}{430} & \multicolumn{1}{c|}{46} & \multicolumn{1}{c|}{354} & \multicolumn{1}{c|}{400} & Design\_27 & 512 & \multicolumn{1}{c|}{94} & \multicolumn{1}{c|}{355} & 449 & \multicolumn{1}{c|}{37} & \multicolumn{1}{c|}{345} & 382 \\ \hline
\multicolumn{1}{|c|}{Design\_136} & \multicolumn{1}{c|}{552} & \multicolumn{1}{c|}{87} & \multicolumn{1}{c|}{387} & \multicolumn{1}{c|}{474} & \multicolumn{1}{c|}{35} & \multicolumn{1}{c|}{410} & \multicolumn{1}{c|}{445} & Design\_30 & 491 & \multicolumn{1}{c|}{86} & \multicolumn{1}{c|}{361} & 447 & \multicolumn{1}{c|}{45} & \multicolumn{1}{c|}{359} & 404 \\ \hline
\multicolumn{1}{|c|}{Design\_137} & \multicolumn{1}{c|}{656} & \multicolumn{1}{c|}{92} & \multicolumn{1}{c|}{380} & \multicolumn{1}{c|}{472} & \multicolumn{1}{c|}{38} & \multicolumn{1}{c|}{401} & \multicolumn{1}{c|}{439} & Design\_31 & 494 & \multicolumn{1}{c|}{85} & \multicolumn{1}{c|}{371} & 456 & \multicolumn{1}{c|}{36} & \multicolumn{1}{c|}{364} & 400 \\ \hline
\multicolumn{1}{|c|}{Design\_140} & \multicolumn{1}{c|}{524} & \multicolumn{1}{c|}{198} & \multicolumn{1}{c|}{398} & \multicolumn{1}{c|}{596} & \multicolumn{1}{c|}{46} & \multicolumn{1}{c|}{410} & \multicolumn{1}{c|}{456} & Design\_32 & 510 & \multicolumn{1}{c|}{92} & \multicolumn{1}{c|}{361} & 453 & \multicolumn{1}{c|}{37} & \multicolumn{1}{c|}{363} & 400 \\ \hline
\multicolumn{1}{|c|}{Design\_141} & \multicolumn{1}{c|}{532} & \multicolumn{1}{c|}{203} & \multicolumn{1}{c|}{366} & \multicolumn{1}{c|}{569} & \multicolumn{1}{c|}{68} & \multicolumn{1}{c|}{355} & \multicolumn{1}{c|}{423} & Design\_35 & 524 & \multicolumn{1}{c|}{95} & \multicolumn{1}{c|}{366} & 461 & \multicolumn{1}{c|}{46} & \multicolumn{1}{c|}{413} & 459 \\ \hline
\multicolumn{1}{|c|}{Design\_142} & \multicolumn{1}{c|}{506} & \multicolumn{1}{c|}{95} & \multicolumn{1}{c|}{365} & \multicolumn{1}{c|}{460} & \multicolumn{1}{c|}{38} & \multicolumn{1}{c|}{375} & \multicolumn{1}{c|}{413} & Design\_36 & 514 & \multicolumn{1}{c|}{90} & \multicolumn{1}{c|}{362} & 452 & \multicolumn{1}{c|}{36} & \multicolumn{1}{c|}{403} & 439 \\ \hline
\multicolumn{1}{|c|}{Design\_145} & \multicolumn{1}{c|}{541} & \multicolumn{1}{c|}{97} & \multicolumn{1}{c|}{361} & \multicolumn{1}{c|}{458} & \multicolumn{1}{c|}{47} & \multicolumn{1}{c|}{373} & \multicolumn{1}{c|}{420} & Design\_37 & 530 & \multicolumn{1}{c|}{89} & \multicolumn{1}{c|}{392} & 481 & \multicolumn{1}{c|}{37} & \multicolumn{1}{c|}{411} & 448 \\ \hline
\multicolumn{1}{|c|}{Design\_147} & \multicolumn{1}{c|}{548} & \multicolumn{1}{c|}{91} & \multicolumn{1}{c|}{400} & \multicolumn{1}{c|}{491} & \multicolumn{1}{c|}{37} & \multicolumn{1}{c|}{411} & \multicolumn{1}{c|}{448} & Design\_40 & 546 & \multicolumn{1}{c|}{96} & \multicolumn{1}{c|}{393} & 489 & \multicolumn{1}{c|}{46} & \multicolumn{1}{c|}{398} & 444 \\ \hline
\multicolumn{1}{|c|}{Design\_15} & \multicolumn{1}{c|}{470} & \multicolumn{1}{c|}{93} & \multicolumn{1}{c|}{341} & \multicolumn{1}{c|}{434} & \multicolumn{1}{c|}{43} & \multicolumn{1}{c|}{347} & \multicolumn{1}{c|}{390} & Design\_41 & 505 & \multicolumn{1}{c|}{91} & \multicolumn{1}{c|}{410} & 501 & \multicolumn{1}{c|}{37} & \multicolumn{1}{c|}{422} & 459 \\ \hline
\multicolumn{1}{|c|}{Design\_150} & \multicolumn{1}{c|}{569} & \multicolumn{1}{c|}{96} & \multicolumn{1}{c|}{421} & \multicolumn{1}{c|}{517} & \multicolumn{1}{c|}{47} & \multicolumn{1}{c|}{424} & \multicolumn{1}{c|}{471} & Design\_42 & 522 & \multicolumn{1}{c|}{90} & \multicolumn{1}{c|}{384} & 474 & \multicolumn{1}{c|}{38} & \multicolumn{1}{c|}{420} & 458 \\ \hline
\multicolumn{1}{|c|}{Design\_151} & \multicolumn{1}{c|}{518} & \multicolumn{1}{c|}{91} & \multicolumn{1}{c|}{363} & \multicolumn{1}{c|}{454} & \multicolumn{1}{c|}{37} & \multicolumn{1}{c|}{376} & \multicolumn{1}{c|}{413} & Design\_45 & 533 & \multicolumn{1}{c|}{95} & \multicolumn{1}{c|}{394} & 489 & \multicolumn{1}{c|}{46} & \multicolumn{1}{c|}{437} & 483 \\ \hline
\multicolumn{1}{|c|}{Design\_152} & \multicolumn{1}{c|}{503} & \multicolumn{1}{c|}{90} & \multicolumn{1}{c|}{379} & \multicolumn{1}{c|}{469} & \multicolumn{1}{c|}{38} & \multicolumn{1}{c|}{399} & \multicolumn{1}{c|}{437} & Design\_46 & 574 & \multicolumn{1}{c|}{92} & \multicolumn{1}{c|}{416} & 508 & \multicolumn{1}{c|}{37} & \multicolumn{1}{c|}{409} & 446 \\ \hline
\multicolumn{1}{|c|}{Design\_155} & \multicolumn{1}{c|}{508} & \multicolumn{1}{c|}{91} & \multicolumn{1}{c|}{365} & \multicolumn{1}{c|}{456} & \multicolumn{1}{c|}{46} & \multicolumn{1}{c|}{411} & \multicolumn{1}{c|}{457} & Design\_47 & 577 & \multicolumn{1}{c|}{92} & \multicolumn{1}{c|}{409} & 501 & \multicolumn{1}{c|}{38} & \multicolumn{1}{c|}{388} & 426 \\ \hline
\multicolumn{1}{|c|}{Design\_156} & \multicolumn{1}{c|}{584} & \multicolumn{1}{c|}{96} & \multicolumn{1}{c|}{436} & \multicolumn{1}{c|}{532} & \multicolumn{1}{c|}{37} & \multicolumn{1}{c|}{493} & \multicolumn{1}{c|}{530} & Design\_5 & 442 & \multicolumn{1}{c|}{85} & \multicolumn{1}{c|}{304} & 389 & \multicolumn{1}{c|}{43} & \multicolumn{1}{c|}{321} & 364 \\ \hline
\multicolumn{1}{|c|}{Design\_16} & \multicolumn{1}{c|}{489} & \multicolumn{1}{c|}{92} & \multicolumn{1}{c|}{336} & \multicolumn{1}{c|}{428} & \multicolumn{1}{c|}{34} & \multicolumn{1}{c|}{386} & \multicolumn{1}{c|}{420} & Design\_50 & 573 & \multicolumn{1}{c|}{101} & \multicolumn{1}{c|}{431} & 532 & \multicolumn{1}{c|}{47} & \multicolumn{1}{c|}{405} & 452 \\ \hline
\multicolumn{1}{|c|}{Design\_160} & \multicolumn{1}{c|}{611} & \multicolumn{1}{c|}{95} & \multicolumn{1}{c|}{457} & \multicolumn{1}{c|}{552} & \multicolumn{1}{c|}{46} & \multicolumn{1}{c|}{498} & \multicolumn{1}{c|}{544} & Design\_51 & 548 & \multicolumn{1}{c|}{92} & \multicolumn{1}{c|}{388} & 480 & \multicolumn{1}{c|}{37} & \multicolumn{1}{c|}{387} & 424 \\ \hline
\multicolumn{1}{|c|}{Design\_161} & \multicolumn{1}{c|}{519} & \multicolumn{1}{c|}{93} & \multicolumn{1}{c|}{375} & \multicolumn{1}{c|}{468} & \multicolumn{1}{c|}{37} & \multicolumn{1}{c|}{423} & \multicolumn{1}{c|}{460} & Design\_52 & 544 & \multicolumn{1}{c|}{90} & \multicolumn{1}{c|}{389} & 479 & \multicolumn{1}{c|}{39} & \multicolumn{1}{c|}{391} & 430 \\ \hline
\multicolumn{1}{|c|}{Design\_162} & \multicolumn{1}{c|}{594} & \multicolumn{1}{c|}{211} & \multicolumn{1}{c|}{387} & \multicolumn{1}{c|}{598} & \multicolumn{1}{c|}{73} & \multicolumn{1}{c|}{436} & \multicolumn{1}{c|}{509} & Design\_55 & 558 & \multicolumn{1}{c|}{95} & \multicolumn{1}{c|}{402} & 497 & \multicolumn{1}{c|}{48} & \multicolumn{1}{c|}{398} & 446 \\ \hline
\multicolumn{1}{|c|}{Design\_165} & \multicolumn{1}{c|}{531} & \multicolumn{1}{c|}{91} & \multicolumn{1}{c|}{382} & \multicolumn{1}{c|}{473} & \multicolumn{1}{c|}{48} & \multicolumn{1}{c|}{439} & \multicolumn{1}{c|}{487} & Design\_56 & 545 & \multicolumn{1}{c|}{94} & \multicolumn{1}{c|}{427} & 521 & \multicolumn{1}{c|}{37} & \multicolumn{1}{c|}{399} & 436 \\ \hline
\multicolumn{1}{|c|}{Design\_166} & \multicolumn{1}{c|}{582} & \multicolumn{1}{c|}{200} & \multicolumn{1}{c|}{436} & \multicolumn{1}{c|}{636} & \multicolumn{1}{c|}{70} & \multicolumn{1}{c|}{487} & \multicolumn{1}{c|}{557} & Design\_57 & 633 & \multicolumn{1}{c|}{96} & \multicolumn{1}{c|}{470} & 566 & \multicolumn{1}{c|}{39} & \multicolumn{1}{c|}{432} & 471 \\ \hline
\multicolumn{1}{|c|}{Design\_167} & \multicolumn{1}{c|}{614} & \multicolumn{1}{c|}{92} & \multicolumn{1}{c|}{445} & \multicolumn{1}{c|}{537} & \multicolumn{1}{c|}{39} & \multicolumn{1}{c|}{492} & \multicolumn{1}{c|}{531} & Design\_6 & 482 & \multicolumn{1}{c|}{83} & \multicolumn{1}{c|}{321} & 404 & \multicolumn{1}{c|}{33} & \multicolumn{1}{c|}{297} & 330 \\ \hline
\multicolumn{1}{|c|}{Design\_17} & \multicolumn{1}{c|}{483} & \multicolumn{1}{c|}{89} & \multicolumn{1}{c|}{334} & \multicolumn{1}{c|}{423} & \multicolumn{1}{c|}{36} & \multicolumn{1}{c|}{406} & \multicolumn{1}{c|}{442} & Design\_60 & 603 & \multicolumn{1}{c|}{98} & \multicolumn{1}{c|}{457} & 555 & \multicolumn{1}{c|}{46} & \multicolumn{1}{c|}{423} & 469 \\ \hline
\multicolumn{1}{|c|}{Design\_170} & \multicolumn{1}{c|}{608} & \multicolumn{1}{c|}{92} & \multicolumn{1}{c|}{460} & \multicolumn{1}{c|}{552} & \multicolumn{1}{c|}{48} & \multicolumn{1}{c|}{503} & \multicolumn{1}{c|}{551} & Design\_61 & 439 & \multicolumn{1}{c|}{84} & \multicolumn{1}{c|}{330} & 414 & \multicolumn{1}{c|}{35} & \multicolumn{1}{c|}{314} & 349 \\ \hline
\multicolumn{1}{|c|}{Design\_171} & \multicolumn{1}{c|}{562} & \multicolumn{1}{c|}{89} & \multicolumn{1}{c|}{395} & \multicolumn{1}{c|}{484} & \multicolumn{1}{c|}{38} & \multicolumn{1}{c|}{447} & \multicolumn{1}{c|}{485} & Design\_62 & 439 & \multicolumn{1}{c|}{84} & \multicolumn{1}{c|}{325} & 409 & \multicolumn{1}{c|}{35} & \multicolumn{1}{c|}{317} & 352 \\ \hline
\multicolumn{1}{|c|}{Design\_172} & \multicolumn{1}{c|}{574} & \multicolumn{1}{c|}{93} & \multicolumn{1}{c|}{404} & \multicolumn{1}{c|}{497} & \multicolumn{1}{c|}{40} & \multicolumn{1}{c|}{446} & \multicolumn{1}{c|}{486} & Design\_65 & 460 & \multicolumn{1}{c|}{85} & \multicolumn{1}{c|}{336} & 421 & \multicolumn{1}{c|}{43} & \multicolumn{1}{c|}{325} & 368 \\ \hline
\multicolumn{1}{|c|}{Design\_175} & \multicolumn{1}{c|}{567} & \multicolumn{1}{c|}{95} & \multicolumn{1}{c|}{395} & \multicolumn{1}{c|}{490} & \multicolumn{1}{c|}{49} & \multicolumn{1}{c|}{446} & \multicolumn{1}{c|}{495} & Design\_66 & 464 & \multicolumn{1}{c|}{81} & \multicolumn{1}{c|}{325} & 406 & \multicolumn{1}{c|}{33} & \multicolumn{1}{c|}{317} & 350 \\ \hline
\multicolumn{1}{|c|}{Design\_176} & \multicolumn{1}{c|}{645} & \multicolumn{1}{c|}{92} & \multicolumn{1}{c|}{465} & \multicolumn{1}{c|}{557} & \multicolumn{1}{c|}{38} & \multicolumn{1}{c|}{508} & \multicolumn{1}{c|}{546} & Design\_67 & 461 & \multicolumn{1}{c|}{84} & \multicolumn{1}{c|}{329} & 413 & \multicolumn{1}{c|}{35} & \multicolumn{1}{c|}{323} & 358 \\ \hline
\multicolumn{1}{|c|}{Design\_180} & \multicolumn{1}{c|}{638} & \multicolumn{1}{c|}{203} & \multicolumn{1}{c|}{478} & \multicolumn{1}{c|}{681} & \multicolumn{1}{c|}{96} & \multicolumn{1}{c|}{549} & \multicolumn{1}{c|}{645} & Design\_7 & 457 & \multicolumn{1}{c|}{85} & \multicolumn{1}{c|}{325} & 410 & \multicolumn{1}{c|}{35} & \multicolumn{1}{c|}{302} & 337 \\ \hline
\multicolumn{1}{|c|}{Design\_181} & \multicolumn{1}{c|}{427} & \multicolumn{1}{c|}{85} & \multicolumn{1}{c|}{292} & \multicolumn{1}{c|}{377} & \multicolumn{1}{c|}{33} & \multicolumn{1}{c|}{353} & \multicolumn{1}{c|}{386} & Design\_70 & 463 & \multicolumn{1}{c|}{87} & \multicolumn{1}{c|}{355} & 442 & \multicolumn{1}{c|}{44} & \multicolumn{1}{c|}{339} & 383 \\ \hline
\multicolumn{1}{|c|}{Design\_182} & \multicolumn{1}{c|}{445} & \multicolumn{1}{c|}{86} & \multicolumn{1}{c|}{299} & \multicolumn{1}{c|}{385} & \multicolumn{1}{c|}{34} & \multicolumn{1}{c|}{354} & \multicolumn{1}{c|}{388} & Design\_71 & 467 & \multicolumn{1}{c|}{84} & \multicolumn{1}{c|}{344} & 428 & \multicolumn{1}{c|}{35} & \multicolumn{1}{c|}{334} & 369 \\ \hline
\multicolumn{1}{|c|}{Design\_185} & \multicolumn{1}{c|}{464} & \multicolumn{1}{c|}{88} & \multicolumn{1}{c|}{307} & \multicolumn{1}{c|}{395} & \multicolumn{1}{c|}{43} & \multicolumn{1}{c|}{358} & \multicolumn{1}{c|}{401} & Design\_72 & 498 & \multicolumn{1}{c|}{86} & \multicolumn{1}{c|}{343} & 429 & \multicolumn{1}{c|}{36} & \multicolumn{1}{c|}{339} & 375 \\ \hline
\multicolumn{1}{|c|}{Design\_186} & \multicolumn{1}{c|}{465} & \multicolumn{1}{c|}{87} & \multicolumn{1}{c|}{293} & \multicolumn{1}{c|}{380} & \multicolumn{1}{c|}{33} & \multicolumn{1}{c|}{354} & \multicolumn{1}{c|}{387} & Design\_75 & 478 & \multicolumn{1}{c|}{87} & \multicolumn{1}{c|}{355} & 442 & \multicolumn{1}{c|}{44} & \multicolumn{1}{c|}{334} & 378 \\ \hline
\multicolumn{1}{|c|}{Design\_187} & \multicolumn{1}{c|}{507} & \multicolumn{1}{c|}{88} & \multicolumn{1}{c|}{296} & \multicolumn{1}{c|}{384} & \multicolumn{1}{c|}{35} & \multicolumn{1}{c|}{302} & \multicolumn{1}{c|}{337} & Design\_76 & 531 & \multicolumn{1}{c|}{84} & \multicolumn{1}{c|}{342} & 426 & \multicolumn{1}{c|}{35} & \multicolumn{1}{c|}{381} & 416 \\ \hline
\multicolumn{1}{|c|}{Design\_190} & \multicolumn{1}{c|}{495} & \multicolumn{1}{c|}{92} & \multicolumn{1}{c|}{320} & \multicolumn{1}{c|}{412} & \multicolumn{1}{c|}{43} & \multicolumn{1}{c|}{348} & \multicolumn{1}{c|}{391} & Design\_77 & 502 & \multicolumn{1}{c|}{84} & \multicolumn{1}{c|}{353} & 437 & \multicolumn{1}{c|}{36} & \multicolumn{1}{c|}{356} & 392 \\ \hline
\multicolumn{1}{|c|}{Design\_191} & \multicolumn{1}{c|}{456} & \multicolumn{1}{c|}{87} & \multicolumn{1}{c|}{297} & \multicolumn{1}{c|}{384} & \multicolumn{1}{c|}{34} & \multicolumn{1}{c|}{339} & \multicolumn{1}{c|}{373} & Design\_80 & 504 & \multicolumn{1}{c|}{87} & \multicolumn{1}{c|}{345} & 432 & \multicolumn{1}{c|}{44} & \multicolumn{1}{c|}{338} & 382 \\ \hline
\multicolumn{1}{|c|}{Design\_192} & \multicolumn{1}{c|}{453} & \multicolumn{1}{c|}{91} & \multicolumn{1}{c|}{312} & \multicolumn{1}{c|}{403} & \multicolumn{1}{c|}{35} & \multicolumn{1}{c|}{347} & \multicolumn{1}{c|}{382} & Design\_81 & 481 & \multicolumn{1}{c|}{83} & \multicolumn{1}{c|}{369} & 452 & \multicolumn{1}{c|}{34} & \multicolumn{1}{c|}{343} & 377 \\ \hline
\multicolumn{1}{|c|}{Design\_195} & \multicolumn{1}{c|}{483} & \multicolumn{1}{c|}{94} & \multicolumn{1}{c|}{324} & \multicolumn{1}{c|}{418} & \multicolumn{1}{c|}{43} & \multicolumn{1}{c|}{342} & \multicolumn{1}{c|}{385} & Design\_82 & 472 & \multicolumn{1}{c|}{85} & \multicolumn{1}{c|}{357} & 442 & \multicolumn{1}{c|}{36} & \multicolumn{1}{c|}{349} & 385 \\ \hline
\multicolumn{1}{|c|}{Design\_196} & \multicolumn{1}{c|}{466} & \multicolumn{1}{c|}{88} & \multicolumn{1}{c|}{310} & \multicolumn{1}{c|}{398} & \multicolumn{1}{c|}{34} & \multicolumn{1}{c|}{311} & \multicolumn{1}{c|}{345} & Design\_85 & 523 & \multicolumn{1}{c|}{92} & \multicolumn{1}{c|}{357} & 449 & \multicolumn{1}{c|}{46} & \multicolumn{1}{c|}{357} & 403 \\ \hline
\multicolumn{1}{|c|}{Design\_197} & \multicolumn{1}{c|}{485} & \multicolumn{1}{c|}{90} & \multicolumn{1}{c|}{333} & \multicolumn{1}{c|}{423} & \multicolumn{1}{c|}{36} & \multicolumn{1}{c|}{335} & \multicolumn{1}{c|}{371} & Design\_86 & 547 & \multicolumn{1}{c|}{85} & \multicolumn{1}{c|}{382} & 467 & \multicolumn{1}{c|}{35} & \multicolumn{1}{c|}{401} & 436 \\ \hline
\multicolumn{1}{|c|}{Design\_2} & \multicolumn{1}{c|}{452} & \multicolumn{1}{c|}{85} & \multicolumn{1}{c|}{314} & \multicolumn{1}{c|}{399} & \multicolumn{1}{c|}{35} & \multicolumn{1}{c|}{299} & \multicolumn{1}{c|}{334} & Design\_87 & 539 & \multicolumn{1}{c|}{86} & \multicolumn{1}{c|}{391} & 477 & \multicolumn{1}{c|}{36} & \multicolumn{1}{c|}{402} & 438 \\ \hline
\multicolumn{1}{|c|}{Design\_20} & \multicolumn{1}{c|}{495} & \multicolumn{1}{c|}{88} & \multicolumn{1}{c|}{322} & \multicolumn{1}{c|}{410} & \multicolumn{1}{c|}{45} & \multicolumn{1}{c|}{347} & \multicolumn{1}{c|}{392} & Design\_90 & 534 & \multicolumn{1}{c|}{86} & \multicolumn{1}{c|}{394} & 480 & \multicolumn{1}{c|}{45} & \multicolumn{1}{c|}{378} & 423 \\ \hline
\multicolumn{1}{|c|}{Design\_200} & \multicolumn{1}{c|}{477} & \multicolumn{1}{c|}{94} & \multicolumn{1}{c|}{331} & \multicolumn{1}{c|}{425} & \multicolumn{1}{c|}{44} & \multicolumn{1}{c|}{331} & \multicolumn{1}{c|}{375} & Design\_91 & 521 & \multicolumn{1}{c|}{88} & \multicolumn{1}{c|}{362} & 450 & \multicolumn{1}{c|}{36} & \multicolumn{1}{c|}{355} & 391 \\ \hline
\multicolumn{1}{|c|}{Design\_201} & \multicolumn{1}{c|}{442} & \multicolumn{1}{c|}{92} & \multicolumn{1}{c|}{354} & \multicolumn{1}{c|}{446} & \multicolumn{1}{c|}{35} & \multicolumn{1}{c|}{345} & \multicolumn{1}{c|}{380} & Design\_92 & 501 & \multicolumn{1}{c|}{87} & \multicolumn{1}{c|}{386} & 473 & \multicolumn{1}{c|}{38} & \multicolumn{1}{c|}{363} & 401 \\ \hline
\multicolumn{1}{|c|}{Design\_202} & \multicolumn{1}{c|}{458} & \multicolumn{1}{c|}{91} & \multicolumn{1}{c|}{318} & \multicolumn{1}{c|}{409} & \multicolumn{1}{c|}{36} & \multicolumn{1}{c|}{352} & \multicolumn{1}{c|}{388} & Design\_95 & 521 & \multicolumn{1}{c|}{92} & \multicolumn{1}{c|}{369} & 461 & \multicolumn{1}{c|}{46} & \multicolumn{1}{c|}{370} & 416 \\ \hline
\multicolumn{1}{|c|}{Design\_205} & \multicolumn{1}{c|}{489} & \multicolumn{1}{c|}{95} & \multicolumn{1}{c|}{328} & \multicolumn{1}{c|}{423} & \multicolumn{1}{c|}{45} & \multicolumn{1}{c|}{346} & \multicolumn{1}{c|}{391} & Design\_96 & 542 & \multicolumn{1}{c|}{85} & \multicolumn{1}{c|}{387} & 472 & \multicolumn{1}{c|}{36} & \multicolumn{1}{c|}{395} & 431 \\ \hline
\multicolumn{1}{|c|}{Design\_206} & \multicolumn{1}{c|}{511} & \multicolumn{1}{c|}{91} & \multicolumn{1}{c|}{324} & \multicolumn{1}{c|}{415} & \multicolumn{1}{c|}{35} & \multicolumn{1}{c|}{325} & \multicolumn{1}{c|}{360} & Design\_97 & 493 & \multicolumn{1}{c|}{86} & \multicolumn{1}{c|}{403} & 489 & \multicolumn{1}{c|}{38} & \multicolumn{1}{c|}{592} & 630 \\ \hline
\hline
\multicolumn{9}{|c|}{\textbf{Overall Geomean Ratio}} & \multicolumn{1}{l|}{1.0000} & \multicolumn{3}{r|}{0.8892} & \multicolumn{3}{r|}{\textbf{0.8036}} \\ \hline
\end{tabular}%
}
\end{table*}

\renewcommand{\arraystretch}{1.0}% Tighter
\begin{table*}[]
\centering
\caption{Comparison of Routability Scores}
\label{tab:rout_comp}
\vspace{-0.1in}
\resizebox{0.95\textwidth}{!}{%
\begin{tabular}{|c|ccc|ccc|c|ccc|ccc|ccccccc}
\hline
\multirow{2}{*}{\textbf{Design}} & \multicolumn{3}{c|}{\textbf{Vivado}} & \multicolumn{3}{c|}{\textbf{MP + Vivado}} & \multirow{2}{*}{\textbf{Design}} & \multicolumn{3}{c|}{\textbf{Vivado}} & \multicolumn{3}{c|}{\textbf{MP + Vivado}} & \multicolumn{1}{c|}{\multirow{2}{*}{\textbf{Design}}} & \multicolumn{3}{c|}{\textbf{Vivado}} & \multicolumn{3}{c|}{\textbf{MP + Vivado}} \\ \cline{2-7} \cline{9-14} \cline{16-21} 
 & \multicolumn{1}{c|}{\textbf{$Sr_i$}} & \multicolumn{1}{c|}{\textbf{$Sr_f$}} & \textbf{$\rho$} & \multicolumn{1}{c|}{\textbf{$Sr_i$}} & \multicolumn{1}{c|}{\textbf{$Sr_f$}} & \textbf{$\rho$} &  & \multicolumn{1}{c|}{\textbf{$Sr_i$}} & \multicolumn{1}{c|}{\textbf{$Sr_f$}} & \textbf{$\rho$} & \multicolumn{1}{c|}{\textbf{$Sr_i$}} & \multicolumn{1}{c|}{\textbf{$Sr_f$}} & \textbf{$\rho$} & \multicolumn{1}{c|}{} & \multicolumn{1}{c|}{\textbf{$Sr_i$}} & \multicolumn{1}{c|}{\textbf{$Sr_f$}} & \multicolumn{1}{c|}{\textbf{$\rho$}} & \multicolumn{1}{c|}{\textbf{$Sr_i$}} & \multicolumn{1}{c|}{\textbf{$Sr_f$}} & \multicolumn{1}{c|}{\textbf{$\rho$}} \\ \hline
 \hline 
Design\_10 & \multicolumn{1}{c|}{1} & \multicolumn{1}{c|}{6} & \textbf{7} & \multicolumn{1}{c|}{2} & \multicolumn{1}{c|}{6} & 8 & Design\_171 & \multicolumn{1}{c|}{2} & \multicolumn{1}{c|}{8} & 10 & \multicolumn{1}{c|}{1} & \multicolumn{1}{c|}{7} & \textbf{8} & \multicolumn{1}{c|}{Design\_27} & \multicolumn{1}{c|}{1} & \multicolumn{1}{c|}{7} & \multicolumn{1}{c|}{8} & \multicolumn{1}{c|}{1} & \multicolumn{1}{c|}{7} & \multicolumn{1}{c|}{8} \\ \hline
Design\_100 & \multicolumn{1}{c|}{5} & \multicolumn{1}{c|}{8} & 13 & \multicolumn{1}{c|}{2} & \multicolumn{1}{c|}{7} & \textbf{9} & Design\_172 & \multicolumn{1}{c|}{1} & \multicolumn{1}{c|}{8} & 9 & \multicolumn{1}{c|}{1} & \multicolumn{1}{c|}{7} & \textbf{8} & \multicolumn{1}{c|}{Design\_30} & \multicolumn{1}{c|}{1} & \multicolumn{1}{c|}{6} & \multicolumn{1}{c|}{7} & \multicolumn{1}{c|}{1} & \multicolumn{1}{c|}{6} & \multicolumn{1}{c|}{7} \\ \hline
Design\_101 & \multicolumn{1}{c|}{1} & \multicolumn{1}{c|}{6} & 7 & \multicolumn{1}{c|}{1} & \multicolumn{1}{c|}{6} & 7 & Design\_175 & \multicolumn{1}{c|}{1} & \multicolumn{1}{c|}{9} & 10 & \multicolumn{1}{c|}{1} & \multicolumn{1}{c|}{7} & \textbf{8} & \multicolumn{1}{c|}{Design\_31} & \multicolumn{1}{c|}{1} & \multicolumn{1}{c|}{7} & \multicolumn{1}{c|}{8} & \multicolumn{1}{c|}{1} & \multicolumn{1}{c|}{5} & \multicolumn{1}{c|}{\textbf{6}} \\ \hline
Design\_102 & \multicolumn{1}{c|}{6} & \multicolumn{1}{c|}{9} & 15 & \multicolumn{1}{c|}{1} & \multicolumn{1}{c|}{7} & \textbf{8} & Design\_176 & \multicolumn{1}{c|}{27} & \multicolumn{1}{c|}{27} & \textbf{54} & \multicolumn{1}{c|}{23} & \multicolumn{1}{c|}{35} & 58 & \multicolumn{1}{c|}{Design\_32} & \multicolumn{1}{c|}{1} & \multicolumn{1}{c|}{6} & \multicolumn{1}{c|}{7} & \multicolumn{1}{c|}{1} & \multicolumn{1}{c|}{6} & \multicolumn{1}{c|}{7} \\ \hline
Design\_105 & \multicolumn{1}{c|}{1} & \multicolumn{1}{c|}{7} & \textbf{8} & \multicolumn{1}{c|}{1} & \multicolumn{1}{c|}{8} & 9 & Design\_180 & \multicolumn{1}{c|}{7} & \multicolumn{1}{c|}{27} & \textbf{34} & \multicolumn{1}{c|}{14} & \multicolumn{1}{c|}{45} & 59 & \multicolumn{1}{c|}{Design\_35} & \multicolumn{1}{c|}{1} & \multicolumn{1}{c|}{5} & \multicolumn{1}{c|}{\textbf{6}} & \multicolumn{1}{c|}{1} & \multicolumn{1}{c|}{6} & \multicolumn{1}{c|}{7} \\ \hline
Design\_106 & \multicolumn{1}{c|}{2} & \multicolumn{1}{c|}{12} & 14 & \multicolumn{1}{c|}{2} & \multicolumn{1}{c|}{7} & \textbf{9} & Design\_181 & \multicolumn{1}{c|}{1} & \multicolumn{1}{c|}{5} & 6 & \multicolumn{1}{c|}{1} & \multicolumn{1}{c|}{5} & 6 & \multicolumn{1}{c|}{Design\_36} & \multicolumn{1}{c|}{1} & \multicolumn{1}{c|}{7} & \multicolumn{1}{c|}{8} & \multicolumn{1}{c|}{1} & \multicolumn{1}{c|}{6} & \multicolumn{1}{c|}{\textbf{7}} \\ \hline
Design\_107 & \multicolumn{1}{c|}{3} & \multicolumn{1}{c|}{13} & 16 & \multicolumn{1}{c|}{2} & \multicolumn{1}{c|}{9} & \textbf{11} & Design\_182 & \multicolumn{1}{c|}{1} & \multicolumn{1}{c|}{5} & 6 & \multicolumn{1}{c|}{1} & \multicolumn{1}{c|}{5} & 6 & \multicolumn{1}{c|}{Design\_37} & \multicolumn{1}{c|}{1} & \multicolumn{1}{c|}{9} & \multicolumn{1}{c|}{\textbf{10}} & \multicolumn{1}{c|}{9} & \multicolumn{1}{c|}{9} & \multicolumn{1}{c|}{18} \\ \hline
Design\_11 & \multicolumn{1}{c|}{1} & \multicolumn{1}{c|}{5} & \textbf{6} & \multicolumn{1}{c|}{1} & \multicolumn{1}{c|}{6} & 7 & Design\_185 & \multicolumn{1}{c|}{1} & \multicolumn{1}{c|}{5} & \textbf{6} & \multicolumn{1}{c|}{1} & \multicolumn{1}{c|}{5} & 6 & \multicolumn{1}{c|}{Design\_40} & \multicolumn{1}{c|}{1} & \multicolumn{1}{c|}{6} & \multicolumn{1}{c|}{\textbf{7}} & \multicolumn{1}{c|}{2} & \multicolumn{1}{c|}{6} & \multicolumn{1}{c|}{8} \\ \hline
Design\_110 & \multicolumn{1}{c|}{5} & \multicolumn{1}{c|}{11} & \textbf{16} & \multicolumn{1}{c|}{7} & \multicolumn{1}{c|}{12} & 19 & Design\_186 & \multicolumn{1}{c|}{4} & \multicolumn{1}{c|}{8} & 12 & \multicolumn{1}{c|}{1} & \multicolumn{1}{c|}{8} & \textbf{9} & \multicolumn{1}{c|}{Design\_41} & \multicolumn{1}{c|}{1} & \multicolumn{1}{c|}{8} & \multicolumn{1}{c|}{9} & \multicolumn{1}{c|}{1} & \multicolumn{1}{c|}{8} & \multicolumn{1}{c|}{9} \\ \hline
Design\_111 & \multicolumn{1}{c|}{1} & \multicolumn{1}{c|}{9} & 10 & \multicolumn{1}{c|}{1} & \multicolumn{1}{c|}{8} & \textbf{9} & Design\_187 & \multicolumn{1}{c|}{2} & \multicolumn{1}{c|}{7} & \textbf{9} & \multicolumn{1}{c|}{7} & \multicolumn{1}{c|}{9} & 16 & \multicolumn{1}{c|}{Design\_42} & \multicolumn{1}{c|}{1} & \multicolumn{1}{c|}{6} & \multicolumn{1}{c|}{7} & \multicolumn{1}{c|}{1} & \multicolumn{1}{c|}{\textbf{7}} & \multicolumn{1}{c|}{8} \\ \hline
Design\_112 & \multicolumn{1}{c|}{1} & \multicolumn{1}{c|}{9} & 10 & \multicolumn{1}{c|}{1} & \multicolumn{1}{c|}{6} & \textbf{7} & Design\_190 & \multicolumn{1}{c|}{15} & \multicolumn{1}{c|}{10} & \textbf{25} & \multicolumn{1}{c|}{23} & \multicolumn{1}{c|}{14} & 37 & \multicolumn{1}{c|}{Design\_45} & \multicolumn{1}{c|}{1} & \multicolumn{1}{c|}{7} & \multicolumn{1}{c|}{8} & \multicolumn{1}{c|}{1} & \multicolumn{1}{c|}{7} & \multicolumn{1}{c|}{8} \\ \hline
Design\_115 & \multicolumn{1}{c|}{1} & \multicolumn{1}{c|}{8} & 9 & \multicolumn{1}{c|}{1} & \multicolumn{1}{c|}{8} & 9 & Design\_191 & \multicolumn{1}{c|}{1} & \multicolumn{1}{c|}{5} & 6 & \multicolumn{1}{c|}{1} & \multicolumn{1}{c|}{5} & 6 & \multicolumn{1}{c|}{Design\_46} & \multicolumn{1}{c|}{1} & \multicolumn{1}{c|}{8} & \multicolumn{1}{c|}{\textbf{9}} & \multicolumn{1}{c|}{2} & \multicolumn{1}{c|}{10} & \multicolumn{1}{c|}{12} \\ \hline
Design\_116 & \multicolumn{1}{c|}{7} & \multicolumn{1}{c|}{12} & 19 & \multicolumn{1}{c|}{7} & \multicolumn{1}{c|}{11} & \textbf{18} & Design\_192 & \multicolumn{1}{c|}{1} & \multicolumn{1}{c|}{6} & \textbf{7} & \multicolumn{1}{c|}{1} & \multicolumn{1}{c|}{7} & 8 & \multicolumn{1}{c|}{Design\_47} & \multicolumn{1}{c|}{2} & \multicolumn{1}{c|}{8} & \multicolumn{1}{c|}{10} & \multicolumn{1}{c|}{2} & \multicolumn{1}{c|}{8} & \multicolumn{1}{c|}{10} \\ \hline
Design\_117 & \multicolumn{1}{c|}{2} & \multicolumn{1}{c|}{15} & 17 & \multicolumn{1}{c|}{4} & \multicolumn{1}{c|}{9} & \textbf{13} & Design\_195 & \multicolumn{1}{c|}{1} & \multicolumn{1}{c|}{6} & 7 & \multicolumn{1}{c|}{1} & \multicolumn{1}{c|}{5} & \textbf{6} & \multicolumn{1}{c|}{Design\_5} & \multicolumn{1}{c|}{1} & \multicolumn{1}{c|}{6} & \multicolumn{1}{c|}{7} & \multicolumn{1}{c|}{1} & \multicolumn{1}{c|}{5} & \multicolumn{1}{c|}{\textbf{6}} \\ \hline
Design\_12 & \multicolumn{1}{c|}{1} & \multicolumn{1}{c|}{6} & 7 & \multicolumn{1}{c|}{1} & \multicolumn{1}{c|}{6} & 7 & Design\_196 & \multicolumn{1}{c|}{1} & \multicolumn{1}{c|}{8} & 9 & \multicolumn{1}{c|}{1} & \multicolumn{1}{c|}{5} & \textbf{6} & \multicolumn{1}{c|}{Design\_50} & \multicolumn{1}{c|}{1} & \multicolumn{1}{c|}{7} & \multicolumn{1}{c|}{\textbf{8}} & \multicolumn{1}{c|}{1} & \multicolumn{1}{c|}{8} & \multicolumn{1}{c|}{9} \\ \hline
Design\_120 & \multicolumn{1}{c|}{22} & \multicolumn{1}{c|}{23} & 45 & \multicolumn{1}{c|}{3} & \multicolumn{1}{c|}{10} & \textbf{13} & Design\_197 & \multicolumn{1}{c|}{2} & \multicolumn{1}{c|}{7} & \textbf{9} & \multicolumn{1}{c|}{3} & \multicolumn{1}{c|}{9} & 12 & \multicolumn{1}{c|}{Design\_51} & \multicolumn{1}{c|}{1} & \multicolumn{1}{c|}{6} & \multicolumn{1}{c|}{\textbf{7}} & \multicolumn{1}{c|}{1} & \multicolumn{1}{c|}{7} & \multicolumn{1}{c|}{8} \\ \hline
Design\_121 & \multicolumn{1}{c|}{1} & \multicolumn{1}{c|}{7} & \textbf{8} & \multicolumn{1}{c|}{1} & \multicolumn{1}{c|}{9} & 10 & Design\_2 & \multicolumn{1}{c|}{1} & \multicolumn{1}{c|}{5} & 6 & \multicolumn{1}{c|}{1} & \multicolumn{1}{c|}{5} & 6 & \multicolumn{1}{c|}{Design\_52} & \multicolumn{1}{c|}{1} & \multicolumn{1}{c|}{6} & \multicolumn{1}{c|}{7} & \multicolumn{1}{c|}{1} & \multicolumn{1}{c|}{6} & \multicolumn{1}{c|}{7} \\ \hline
Design\_122 & \multicolumn{1}{c|}{1} & \multicolumn{1}{c|}{6} & 7 & \multicolumn{1}{c|}{1} & \multicolumn{1}{c|}{6} & 7 & Design\_20 & \multicolumn{1}{c|}{1} & \multicolumn{1}{c|}{6} & \textbf{7} & \multicolumn{1}{c|}{3} & \multicolumn{1}{c|}{6} & 9 & \multicolumn{1}{c|}{Design\_55} & \multicolumn{1}{c|}{1} & \multicolumn{1}{c|}{9} & \multicolumn{1}{c|}{10} & \multicolumn{1}{c|}{1} & \multicolumn{1}{c|}{9} & \multicolumn{1}{c|}{10} \\ \hline
Design\_125 & \multicolumn{1}{c|}{1} & \multicolumn{1}{c|}{6} & \textbf{7} & \multicolumn{1}{c|}{1} & \multicolumn{1}{c|}{7} & 8 & Design\_200 & \multicolumn{1}{c|}{1} & \multicolumn{1}{c|}{6} & \textbf{7} & \multicolumn{1}{c|}{6} & \multicolumn{1}{c|}{7} & 13 & \multicolumn{1}{c|}{Design\_56} & \multicolumn{1}{c|}{1} & \multicolumn{1}{c|}{9} & \multicolumn{1}{c|}{10} & \multicolumn{1}{c|}{2} & \multicolumn{1}{c|}{8} & \multicolumn{1}{c|}{10} \\ \hline
Design\_126 & \multicolumn{1}{c|}{3} & \multicolumn{1}{c|}{8} & 11 & \multicolumn{1}{c|}{2} & \multicolumn{1}{c|}{8} & \textbf{10} & Design\_201 & \multicolumn{1}{c|}{1} & \multicolumn{1}{c|}{7} & 8 & \multicolumn{1}{c|}{1} & \multicolumn{1}{c|}{6} & \textbf{7} & \multicolumn{1}{c|}{Design\_57} & \multicolumn{1}{c|}{4} & \multicolumn{1}{c|}{8} & \multicolumn{1}{c|}{12} & \multicolumn{1}{c|}{3} & \multicolumn{1}{c|}{8} & \multicolumn{1}{c|}{\textbf{11}} \\ \hline
Design\_127 & \multicolumn{1}{c|}{1} & \multicolumn{1}{c|}{7} & \textbf{8} & \multicolumn{1}{c|}{2} & \multicolumn{1}{c|}{8} & 10 & Design\_202 & \multicolumn{1}{c|}{1} & \multicolumn{1}{c|}{6} & 7 & \multicolumn{1}{c|}{1} & \multicolumn{1}{c|}{5} & \textbf{6} & \multicolumn{1}{c|}{Design\_6} & \multicolumn{1}{c|}{1} & \multicolumn{1}{c|}{6} & \multicolumn{1}{c|}{\textbf{7}} & \multicolumn{1}{c|}{1} & \multicolumn{1}{c|}{7} & \multicolumn{1}{c|}{8} \\ \hline
Design\_130 & \multicolumn{1}{c|}{1} & \multicolumn{1}{c|}{7} & \textbf{8} & \multicolumn{1}{c|}{2} & \multicolumn{1}{c|}{7} & 9 & Design\_205 & \multicolumn{1}{c|}{1} & \multicolumn{1}{c|}{5} & 6 & \multicolumn{1}{c|}{1} & \multicolumn{1}{c|}{5} & 6 & \multicolumn{1}{c|}{Design\_60} & \multicolumn{1}{c|}{1} & \multicolumn{1}{c|}{14} & \multicolumn{1}{c|}{15} & \multicolumn{1}{c|}{1} & \multicolumn{1}{c|}{10} & \multicolumn{1}{c|}{\textbf{11}} \\ \hline
Design\_131 & \multicolumn{1}{c|}{2} & \multicolumn{1}{c|}{6} & 8 & \multicolumn{1}{c|}{1} & \multicolumn{1}{c|}{6} & \textbf{7} & Design\_206 & \multicolumn{1}{c|}{6} & \multicolumn{1}{c|}{10} & 16 & \multicolumn{1}{c|}{3} & \multicolumn{1}{c|}{9} & \textbf{12} & \multicolumn{1}{c|}{Design\_61} & \multicolumn{1}{c|}{1} & \multicolumn{1}{c|}{6} & \multicolumn{1}{c|}{7} & \multicolumn{1}{c|}{1} & \multicolumn{1}{c|}{6} & \multicolumn{1}{c|}{7} \\ \hline
Design\_132 & \multicolumn{1}{c|}{1} & \multicolumn{1}{c|}{6} & 7 & \multicolumn{1}{c|}{1} & \multicolumn{1}{c|}{6} & 7 & Design\_207 & \multicolumn{1}{c|}{3} & \multicolumn{1}{c|}{7} & \textbf{10} & \multicolumn{1}{c|}{5} & \multicolumn{1}{c|}{8} & 13 & \multicolumn{1}{c|}{Design\_62} & \multicolumn{1}{c|}{1} & \multicolumn{1}{c|}{6} & \multicolumn{1}{c|}{7} & \multicolumn{1}{c|}{1} & \multicolumn{1}{c|}{6} & \multicolumn{1}{c|}{7} \\ \hline
Design\_135 & \multicolumn{1}{c|}{1} & \multicolumn{1}{c|}{9} & \textbf{10} & \multicolumn{1}{c|}{1} & \multicolumn{1}{c|}{10} & 11 & Design\_21 & \multicolumn{1}{c|}{1} & \multicolumn{1}{c|}{6} & \textbf{7} & \multicolumn{1}{c|}{1} & \multicolumn{1}{c|}{10} & 11 & \multicolumn{1}{c|}{Design\_65} & \multicolumn{1}{c|}{1} & \multicolumn{1}{c|}{6} & \multicolumn{1}{c|}{7} & \multicolumn{1}{c|}{1} & \multicolumn{1}{c|}{6} & \multicolumn{1}{c|}{7} \\ \hline
Design\_136 & \multicolumn{1}{c|}{14} & \multicolumn{1}{c|}{10} & 24 & \multicolumn{1}{c|}{10} & \multicolumn{1}{c|}{11} & \textbf{21} & Design\_210 & \multicolumn{1}{c|}{3} & \multicolumn{1}{c|}{6} & \textbf{9} & \multicolumn{1}{c|}{5} & \multicolumn{1}{c|}{10} & 15 & \multicolumn{1}{c|}{Design\_66} & \multicolumn{1}{c|}{1} & \multicolumn{1}{c|}{6} & \multicolumn{1}{c|}{7} & \multicolumn{1}{c|}{1} & \multicolumn{1}{c|}{6} & \multicolumn{1}{c|}{7} \\ \hline
Design\_137 & \multicolumn{1}{c|}{17} & \multicolumn{1}{c|}{9} & 26 & \multicolumn{1}{c|}{12} & \multicolumn{1}{c|}{10} & \textbf{22} & Design\_211 & \multicolumn{1}{c|}{1} & \multicolumn{1}{c|}{6} & 7 & \multicolumn{1}{c|}{1} & \multicolumn{1}{c|}{6} & 7 & \multicolumn{1}{c|}{Design\_67} & \multicolumn{1}{c|}{2} & \multicolumn{1}{c|}{6} & \multicolumn{1}{c|}{8} & \multicolumn{1}{c|}{1} & \multicolumn{1}{c|}{7} & \multicolumn{1}{c|}{8} \\ \hline
Design\_140 & \multicolumn{1}{c|}{16} & \multicolumn{1}{c|}{13} & 29 & \multicolumn{1}{c|}{11} & \multicolumn{1}{c|}{11} & \textbf{22} & Design\_212 & \multicolumn{1}{c|}{1} & \multicolumn{1}{c|}{6} & 7 & \multicolumn{1}{c|}{1} & \multicolumn{1}{c|}{5} & \textbf{6} & \multicolumn{1}{c|}{Design\_7} & \multicolumn{1}{c|}{1} & \multicolumn{1}{c|}{6} & \multicolumn{1}{c|}{7} & \multicolumn{1}{c|}{2} & \multicolumn{1}{c|}{5} & \multicolumn{1}{c|}{7} \\ \hline
Design\_141 & \multicolumn{1}{c|}{1} & \multicolumn{1}{c|}{6} & \textbf{7} & \multicolumn{1}{c|}{1} & \multicolumn{1}{c|}{8} & 9 & Design\_215 & \multicolumn{1}{c|}{1} & \multicolumn{1}{c|}{5} & 6 & \multicolumn{1}{c|}{1} & \multicolumn{1}{c|}{5} & 6 & \multicolumn{1}{c|}{Design\_70} & \multicolumn{1}{c|}{1} & \multicolumn{1}{c|}{6} & \multicolumn{1}{c|}{\textbf{7}} & \multicolumn{1}{c|}{6} & \multicolumn{1}{c|}{8} & \multicolumn{1}{c|}{14} \\ \hline
Design\_142 & \multicolumn{1}{c|}{1} & \multicolumn{1}{c|}{7} & \textbf{8} & \multicolumn{1}{c|}{1} & \multicolumn{1}{c|}{10} & 11 & Design\_216 & \multicolumn{1}{c|}{5} & \multicolumn{1}{c|}{9} & 14 & \multicolumn{1}{c|}{2} & \multicolumn{1}{c|}{7} & \textbf{9} & \multicolumn{1}{c|}{Design\_71} & \multicolumn{1}{c|}{1} & \multicolumn{1}{c|}{7} & \multicolumn{1}{c|}{8} & \multicolumn{1}{c|}{1} & \multicolumn{1}{c|}{6} & \multicolumn{1}{c|}{\textbf{7}} \\ \hline
Design\_145 & \multicolumn{1}{c|}{1} & \multicolumn{1}{c|}{8} & \textbf{9} & \multicolumn{1}{c|}{2} & \multicolumn{1}{c|}{8} & 10 & Design\_217 & \multicolumn{1}{c|}{2} & \multicolumn{1}{c|}{7} & 9 & \multicolumn{1}{c|}{1} & \multicolumn{1}{c|}{7} & \textbf{8} & \multicolumn{1}{c|}{Design\_72} & \multicolumn{1}{c|}{1} & \multicolumn{1}{c|}{6} & \multicolumn{1}{c|}{\textbf{7}} & \multicolumn{1}{c|}{1} & \multicolumn{1}{c|}{7} & \multicolumn{1}{c|}{8} \\ \hline
Design\_147 & \multicolumn{1}{c|}{3} & \multicolumn{1}{c|}{10} & 13 & \multicolumn{1}{c|}{2} & \multicolumn{1}{c|}{11} & 13 & Design\_22 & \multicolumn{1}{c|}{1} & \multicolumn{1}{c|}{5} & \textbf{6} & \multicolumn{1}{c|}{1} & \multicolumn{1}{c|}{6} & 7 & \multicolumn{1}{c|}{Design\_75} & \multicolumn{1}{c|}{1} & \multicolumn{1}{c|}{8} & \multicolumn{1}{c|}{9} & \multicolumn{1}{c|}{1} & \multicolumn{1}{c|}{6} & \multicolumn{1}{c|}{\textbf{7}} \\ \hline
Design\_15 & \multicolumn{1}{c|}{1} & \multicolumn{1}{c|}{6} & 7 & \multicolumn{1}{c|}{1} & \multicolumn{1}{c|}{5} & \textbf{6} & Design\_220 & \multicolumn{1}{c|}{3} & \multicolumn{1}{c|}{7} & 10 & \multicolumn{1}{c|}{1} & \multicolumn{1}{c|}{9} & 10 & \multicolumn{1}{c|}{Design\_76} & \multicolumn{1}{c|}{1} & \multicolumn{1}{c|}{6} & \multicolumn{1}{c|}{\textbf{7}} & \multicolumn{1}{c|}{1} & \multicolumn{1}{c|}{7} & \multicolumn{1}{c|}{8} \\ \hline
Design\_150 & \multicolumn{1}{c|}{2} & \multicolumn{1}{c|}{10} & 12 & \multicolumn{1}{c|}{1} & \multicolumn{1}{c|}{8} & \textbf{9} & Design\_221 & \multicolumn{1}{c|}{1} & \multicolumn{1}{c|}{10} & 11 & \multicolumn{1}{c|}{1} & \multicolumn{1}{c|}{6} & \textbf{7} & \multicolumn{1}{c|}{Design\_77} & \multicolumn{1}{c|}{1} & \multicolumn{1}{c|}{8} & \multicolumn{1}{c|}{9} & \multicolumn{1}{c|}{1} & \multicolumn{1}{c|}{8} & \multicolumn{1}{c|}{9} \\ \hline
Design\_151 & \multicolumn{1}{c|}{2} & \multicolumn{1}{c|}{7} & 9 & \multicolumn{1}{c|}{1} & \multicolumn{1}{c|}{8} & 9 & Design\_222 & \multicolumn{1}{c|}{1} & \multicolumn{1}{c|}{5} & \textbf{6} & \multicolumn{1}{c|}{1} & \multicolumn{1}{c|}{6} & 7 & \multicolumn{1}{c|}{Design\_80} & \multicolumn{1}{c|}{2} & \multicolumn{1}{c|}{8} & \multicolumn{1}{c|}{10} & \multicolumn{1}{c|}{1} & \multicolumn{1}{c|}{7} & \multicolumn{1}{c|}{\textbf{8}} \\ \hline
Design\_152 & \multicolumn{1}{c|}{1} & \multicolumn{1}{c|}{7} & \textbf{8} & \multicolumn{1}{c|}{1} & \multicolumn{1}{c|}{11} & 12 & Design\_225 & \multicolumn{1}{c|}{1} & \multicolumn{1}{c|}{6} & 7 & \multicolumn{1}{c|}{1} & \multicolumn{1}{c|}{5} & \textbf{6} & \multicolumn{1}{c|}{Design\_81} & \multicolumn{1}{c|}{1} & \multicolumn{1}{c|}{6} & \multicolumn{1}{c|}{\textbf{7}} & \multicolumn{1}{c|}{1} & \multicolumn{1}{c|}{7} & \multicolumn{1}{c|}{8} \\ \hline
Design\_155 & \multicolumn{1}{c|}{1} & \multicolumn{1}{c|}{6} & \textbf{7} & \multicolumn{1}{c|}{1} & \multicolumn{1}{c|}{7} & 8 & Design\_226 & \multicolumn{1}{c|}{2} & \multicolumn{1}{c|}{7} & \textbf{9} & \multicolumn{1}{c|}{3} & \multicolumn{1}{c|}{7} & 10 & \multicolumn{1}{c|}{Design\_82} & \multicolumn{1}{c|}{1} & \multicolumn{1}{c|}{6} & \multicolumn{1}{c|}{7} & \multicolumn{1}{c|}{1} & \multicolumn{1}{c|}{6} & \multicolumn{1}{c|}{7} \\ \hline
Design\_156 & \multicolumn{1}{c|}{6} & \multicolumn{1}{c|}{13} & 19 & \multicolumn{1}{c|}{4} & \multicolumn{1}{c|}{9} & \textbf{13} & Design\_227 & \multicolumn{1}{c|}{7} & \multicolumn{1}{c|}{8} & \textbf{15} & \multicolumn{1}{c|}{6} & \multicolumn{1}{c|}{11} & 17 & \multicolumn{1}{c|}{Design\_85} & \multicolumn{1}{c|}{1} & \multicolumn{1}{c|}{5} & \multicolumn{1}{c|}{\textbf{6}} & \multicolumn{1}{c|}{1} & \multicolumn{1}{c|}{7} & \multicolumn{1}{c|}{8} \\ \hline
Design\_16 & \multicolumn{1}{c|}{1} & \multicolumn{1}{c|}{6} & \textbf{7} & \multicolumn{1}{c|}{1} & \multicolumn{1}{c|}{7} & 8 & Design\_230 & \multicolumn{1}{c|}{3} & \multicolumn{1}{c|}{10} & \textbf{13} & \multicolumn{1}{c|}{8} & \multicolumn{1}{c|}{11} & 19 & \multicolumn{1}{c|}{Design\_86} & \multicolumn{1}{c|}{1} & \multicolumn{1}{c|}{12} & \multicolumn{1}{c|}{13} & \multicolumn{1}{c|}{2} & \multicolumn{1}{c|}{10} & \multicolumn{1}{c|}{\textbf{12}} \\ \hline
Design\_160 & \multicolumn{1}{c|}{3} & \multicolumn{1}{c|}{8} & \textbf{11} & \multicolumn{1}{c|}{8} & \multicolumn{1}{c|}{8} & 16 & Design\_231 & \multicolumn{1}{c|}{1} & \multicolumn{1}{c|}{6} & \textbf{7} & \multicolumn{1}{c|}{1} & \multicolumn{1}{c|}{7} & 8 & \multicolumn{1}{c|}{Design\_87} & \multicolumn{1}{c|}{7} & \multicolumn{1}{c|}{9} & \multicolumn{1}{c|}{\textbf{16}} & \multicolumn{1}{c|}{9} & \multicolumn{1}{c|}{9} & \multicolumn{1}{c|}{18} \\ \hline
Design\_161 & \multicolumn{1}{c|}{1} & \multicolumn{1}{c|}{6} & 7 & \multicolumn{1}{c|}{1} & \multicolumn{1}{c|}{8} & \textbf{9} & Design\_232 & \multicolumn{1}{c|}{1} & \multicolumn{1}{c|}{6} & 7 & \multicolumn{1}{c|}{1} & \multicolumn{1}{c|}{5} & 6 & \multicolumn{1}{c|}{Design\_90} & \multicolumn{1}{c|}{1} & \multicolumn{1}{c|}{7} & \multicolumn{1}{c|}{8} & \multicolumn{1}{c|}{1} & \multicolumn{1}{c|}{8} & \multicolumn{1}{c|}{9} \\ \hline
Design\_162 & \multicolumn{1}{c|}{2} & \multicolumn{1}{c|}{9} & 11 & \multicolumn{1}{c|}{1} & \multicolumn{1}{c|}{8} & \textbf{9} & Design\_235 & \multicolumn{1}{c|}{1} & \multicolumn{1}{c|}{6} & 7 & \multicolumn{1}{c|}{1} & \multicolumn{1}{c|}{6} & 7 & \multicolumn{1}{c|}{Design\_91} & \multicolumn{1}{c|}{1} & \multicolumn{1}{c|}{7} & \multicolumn{1}{c|}{8} & \multicolumn{1}{c|}{1} & \multicolumn{1}{c|}{7} & \multicolumn{1}{c|}{8} \\ \hline
Design\_165 & \multicolumn{1}{c|}{1} & \multicolumn{1}{c|}{8} & 9 & \multicolumn{1}{c|}{1} & \multicolumn{1}{c|}{7} & \textbf{8} & Design\_236 & \multicolumn{1}{c|}{5} & \multicolumn{1}{c|}{9} & \textbf{14} & \multicolumn{1}{c|}{6} & \multicolumn{1}{c|}{9} & 15 & \multicolumn{1}{c|}{Design\_92} & \multicolumn{1}{c|}{2} & \multicolumn{1}{c|}{7} & \multicolumn{1}{c|}{9} & \multicolumn{1}{c|}{1} & \multicolumn{1}{c|}{7} & \multicolumn{1}{c|}{\textbf{8}} \\ \hline
Design\_166 & \multicolumn{1}{c|}{7} & \multicolumn{1}{c|}{11} & 18 & \multicolumn{1}{c|}{5} & \multicolumn{1}{c|}{12} & \textbf{17} & Design\_237 & \multicolumn{1}{c|}{2} & \multicolumn{1}{c|}{7} & \textbf{9} & \multicolumn{1}{c|}{6} & \multicolumn{1}{c|}{8} & 14 & \multicolumn{1}{c|}{Design\_95} & \multicolumn{1}{c|}{1} & \multicolumn{1}{c|}{7} & \multicolumn{1}{c|}{8} & \multicolumn{1}{c|}{1} & \multicolumn{1}{c|}{7} & \multicolumn{1}{c|}{8} \\ \hline
Design\_167 & \multicolumn{1}{c|}{6} & \multicolumn{1}{c|}{18} & 24 & \multicolumn{1}{c|}{7} & \multicolumn{1}{c|}{13} & \textbf{20} & Design\_240 & \multicolumn{1}{c|}{6} & \multicolumn{1}{c|}{8} & 14 & \multicolumn{1}{c|}{4} & \multicolumn{1}{c|}{8} & \textbf{12} & \multicolumn{1}{c|}{Design\_96} & \multicolumn{1}{c|}{1} & \multicolumn{1}{c|}{7} & \multicolumn{1}{c|}{8} & \multicolumn{1}{c|}{1} & \multicolumn{1}{c|}{7} & \multicolumn{1}{c|}{8} \\ \hline
Design\_17 & \multicolumn{1}{c|}{1} & \multicolumn{1}{c|}{6} & \textbf{7} & \multicolumn{1}{c|}{6} & \multicolumn{1}{c|}{11} & 17 & Design\_25 & \multicolumn{1}{c|}{1} & \multicolumn{1}{c|}{6} & 7 & \multicolumn{1}{c|}{1} & \multicolumn{1}{c|}{5} & \textbf{6} & \multicolumn{1}{c|}{Design\_97} & \multicolumn{1}{c|}{1} & \multicolumn{1}{c|}{6} & \multicolumn{1}{c|}{\textbf{7}} & \multicolumn{1}{c|}{1} & \multicolumn{1}{c|}{7} & \multicolumn{1}{c|}{8} \\ \hline
Design\_170 & \multicolumn{1}{c|}{17} & \multicolumn{1}{c|}{37} & 54 & \multicolumn{1}{c|}{19} & \multicolumn{1}{c|}{21} & \textbf{40} & Design\_26 & \multicolumn{1}{c|}{1} & \multicolumn{1}{c|}{7} & 8 & \multicolumn{1}{c|}{1} & \multicolumn{1}{c|}{7} & 8 & \multicolumn{7}{c}{} \\ \cline{1-14}
\end{tabular}%
}
\end{table*}

\subsubsection{Routability}
%%%%%%%%%%%%%%%
In the evaluation of routability, we adopted the same scoring as the contest, outlined in Equation ~\eqref{eqn:cong_score}. 
The macro placement solutions from both CPU and GPU versions of \sys{} are similar and are compared with the baseline Vivado in Table ~\ref{tab:rout_comp}.
The best routability scores are highlighted in \textbf{bold} and it can be observed that both approaches have similar scores, with \sys{} being $1\%$ higher across all the 140 designs.

%% file: texts/Conclusion.tex
\section{Conclusion}\label{sec:conclusion}
This work presents \sys{}, an open-source, GPU-accelerated macro-placer for modern heterogeneous FPGAs that can efficiently handle macros with complex constraints such as cascade shapes and region constraints.
The approaches employed in this work are highly adaptable and can be applied to different FPGA architectures and placement formulations.
\sys{} has the option to perform GPU acceleration with optimized operators for improved scalability.
\sys{} is among the top contestants on the MLCAD 2023 FPGA Macro-Placement Contest~\cite{mlcad2023}.

We plan to employ machine learning techniques to enhance the macro placement quality and refinement schemes in future work.